\documentclass[twocolumn]{aastex61}
\usepackage{devanagari}
\usepackage{color}

\shorttitle{Saraswati: An   extremely  massive supercluster}
\shortauthors{Bagchi et al.}

\begin{document}

\title{ \textbf{Saraswati: An Extremely  Massive $\sim200$~Megaparsec Scale Supercluster}}

\vspace{5cm}

\correspondingauthor{Joydeep Bagchi}
\email{joydeep@iucaa.in}

\author[0000-0002-2922-2884]{Joydeep Bagchi}
\affiliation{The Inter-University Centre for Astronomy and Astrophysics (IUCAA),
S.P. Pune University Campus, Post Bag 4, Pune  411007, India}

\author[0000-0003-2601-2707]{Shishir Sankhyayan}
\affiliation{Indian Institute of Science Education and Research (IISER),
Dr. Homi Bhabha Road, Pashan, Pune 411008, India}

\author[0000-0002-4404-916X]{Prakash Sarkar}
\affiliation{Dept. of Physics, National Institute of Technology (NIT),
NIT Campus, P.O. RIT, Jamshedpur 831014, India}

\author[0000-0002-4864-4046]{Somak Raychaudhury}
\affiliation{The Inter-University Centre for Astronomy and Astrophysics (IUCAA),
S.P. Pune University Campus, Post Bag 4, Pune  411007, India}

\author[0000-0003-0423-5002]{Joe Jacob}
\affiliation{Department of Physics, Newman College, Thodupuzha 685585, India}

\author[0000-0001-9212-3574]{Pratik Dabhade}
\affiliation{The Inter-University Centre for Astronomy and Astrophysics (IUCAA),
S.P. Pune University Campus, Post Bag 4, Pune  411007, India}



\begin{abstract}
Here we report the discovery of an  extremely  massive  and large supercluster  
(called  Saraswati\footnote{Saraswati ({\dn sr-vtF}) is the 
ancient Indian  goddess of knowledge, music, art, wisdom, and nature, muse of
all creative endeavor. Historically, the Saraswati river  is an important river
goddess mentioned in the {\em Rig'veda}. The  river could not be identified with any present day rivers. 
 Saraswati played an important role in Indian culture, since Vedic Sanskrit and the first part of the {\em Rig'veda} are regarded 
to have originated when the Vedic people lived on its banks, during the second millennium BCE.
The Sanskrit name also means ``ever flowing stream with many 
pools." This may describe the present large-scale filamentary structure  of galaxies located in the
Zodiacal constellation of ``Pisces," having  many clusters and groups moving  and merging together. (\url{https://en.wikipedia.org/wiki/Saraswati})})
 found in the Stripe 82 region of SDSS. This supercluster is a 
major  concentration of  galaxies and galaxy clusters, forming a wall-like 
structure spanning at least 200 Mpc across at
the redshift $z \approx 0.3$. This enormous structure  is surrounded by a network of
galaxy filaments, clusters, and  large,   $\sim40 - 170$~Mpc diameter,  voids.
The mean  density contrast $\delta$ (relative to the background  matter density of the universe) of 
Saraswati is $\ga 1.62$ and the main body of the supercluster  comprises  at least 43 
massive galaxy clusters  (mean $z = 0.28$) with a total  mass of
$\sim 2 \times 10^{16} M_{\odot}$. The spherical collapse model suggests that the
central region of radius  $\sim20$ Mpc and mass at least $ 4 \times 10^{15} M_{\odot}$ 
may be collapsing. This  places  it
among the  few  largest and  most  massive superclusters  known, comparable to the 
most massive  `Shapley Concentration' ($z \approx 0.046$)  in the nearby  universe.
The  Saraswati  supercluster   and its environs  
reveal that some  extreme large-scale, prominent matter density enhancements had 
formed  $\sim4$ Gy in the past   when  dark energy had just started to
dominate  structure  formation. This  galactic concentration  sheds  light on    the
role of dark  energy and cosmological initial conditions in supercluster 
formation, and   tests  the  competing cosmological models.

\end{abstract}


\section{Introduction}

In the leading paradigm of Cold Dark Matter ($\Lambda$CDM) cosmology,
large-scale structures  assemble hierarchically through the
gravitational clustering of matter.  An interplay of dark matter and
dark energy results in an  intricate  pattern   of  
interconnected filaments, wall-like pancakes,  and  dense  clusters surrounded by large 
near-empty void regions   \citep{2014dmcw.book.....E}. 
Superclusters of galaxies are believed to be the largest concentrations of matter in the 
universe  whose origin is still being debated 
\citep{1956VA......2.1584D,1961AJ.....66..607A,1978MNRAS.185..357J,1994MNRAS.269..301E}.
Occasionally spanning over a hundred megaparsecs (Mpc), superclusters usually display
a pronounced filamentary or sheet-like morphology, and are surrounded by large almost empty regions
called voids \citep{2014dmcw.book.....E}. Taken together, the  supercluster-void distribution is
commonly referred to as the `Cosmic Web' \citep{1996Natur.380..603B,2014dmcw.book.....E}.
One has been able to associate superclusters with a network  of galaxies  
connected to  a few  rich cluster systems,
as well as with denser and  more massive  structures characterized
by  the concentration  of several very massive clusters.
Contemporary catalogs of superclusters therefore range from relatively smaller systems
such as  the Hercules supercluster \citep{1961AJ.....66..607A} to  the much  more massive
{\em Shapley Concentration} containing several  major clusters and numerous 
groups \citep{1989Natur.342..251R,2000MNRAS.312..540B,Proust06}. More recently,  a new supercluster  
called {\em Laniakea}, spanning   $\sim160$ Mpc across,  is identified \citep{2014Natur.513...71T}
which is home to our Milky Way galaxy. Another  large,  
wall-like  structure  spanning  $\sim250$ Mpc was  discovered in the SDSS-BOSS survey  
(the {\em BOSS Great Wall}) at  redshift 0.47  \citep{2016A&A...588L...4L}.

Unlike clusters,  superclusters   have  not yet   virialized  and  so may
retain some  memory of  their initial conditions that led to their
formation. The assembly of the largest superclusters
is still a  matter of  conjecture, with arguments being given both in favor of assembly through
gravitational clustering \citep{2012ApJ...759L...7P}  and the more radical point of
view that  large  superclusters should be viewed as  rare objects \citep{2011MNRAS.417.2938S}, which might even
challenge the widely held  Copernican principle \citep{2003imcs.book.....L}. Therefore, 
large-scale matter overdensities represented by clusters and superclusters, especially those identified
at early cosmological epochs   $z \gtrsim 0.3$,  (i) are sensitive probes of the
presence of dark matter and dark energy  at higher redshifts, (ii) 
shed light on how  structure  might form in the universe,
and (iii) provide model-independent tests of general relativity on the largest scales by
extending solar system tests by 8--10 orders of magnitude \citep{2010AnPhy.325.1479J}.

The primary focus of  this  paper  is  Saraswati, 
an unusually  massive and  large-scale ($\sim200$~Mpc) 
supercluster of galaxies at a  mean redshift $z = 0.28$, and  its properties, identified in 
the  Sloan Digital Sky Survey (SDSS-III) data. This wall-like 
supercluster is  shown to  be  associated with at least 43 massive galaxy clusters and groups,
of which  a  few   extremely  massive  ones  constitute its  bound,  central  core region.  
Our observations show  the  Saraswati to be  highly unusual in terms of 
its  morphology, mass, and richness, and  in  detection of  localized  non-thermal 
radio emission,  relating to its complex dynamical 
state.  We have  highlighted the importance of our findings for  ongoing  cosmological studies
of dark energy  and the  growth of the large-scale   structures in the universe.
We  have used the  following  cosmological parameters  based on five-year 
WMAP results \citep{2009ApJS..180..330K};  $\Omega_M = 0.279$, $\Omega_\Lambda = 0.721$,
$\Omega_R = 8.493 \times 10^{-5}$, $\omega_0 = -1$, and
$H_0 = 70.1$ km s$^{-1}$ Mpc$^{-1}$,  which results  in  scale factor of 4.47 
kpc arcsec$^{-1}$  for a redshift of 0.3. 
Below $M_{n}$ and $R_{n}$ denote the total mass and radius corresponding to 
a total density contrast $n \rho(z)$, where $\rho(z)$ is the critical density 
of  the universe at redshift $z$. 

\section{Identifying an  Exceptionally Massive and Large-scale Supercluster}

\begin{figure}
\centering
\includegraphics[scale=0.766]{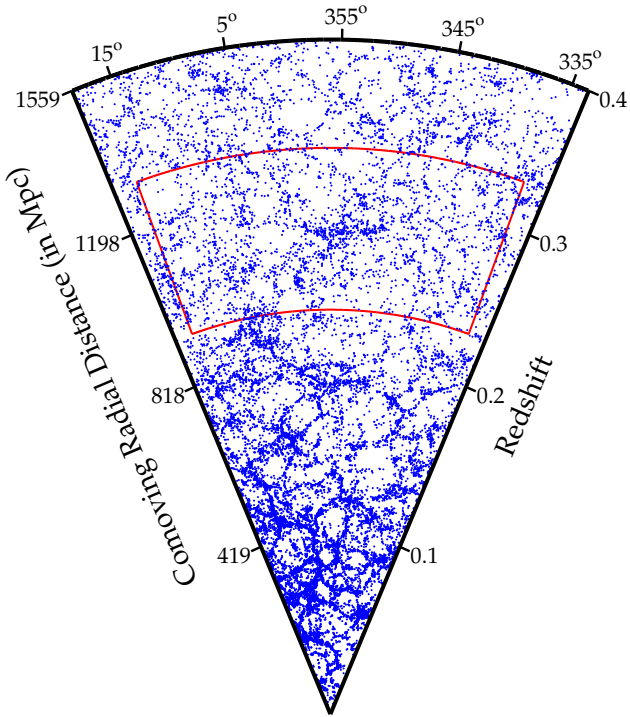}
\caption{Redshift cone plot of SDSS  galaxies is shown (LBS sample; see text), 
extending from redshift $z =$ 0 to 0.4,  and  R.A. width $45 \deg$,
decl. width $2.5 \deg $,   centered near  Saraswati supercluster, the
prominent large-scale structure  within the red box. The small  depth in  decl.  
direction has been suppressed for clarity. \label{fig:zcone}}
\end{figure}

Our   primary results, pertaining to the discovery of a $\sim200$ Mpc scale, 
 massive supercluster (Saraswati) in the  distribution of  galaxies is  
 shown in Figures~\ref{fig:zcone}, \ref{fig:LOWZclust}, \ref{fig:LBS_clust}, and 
 the  subsequent figures. A  redshift cone plot of  SDSS  galaxies,  between $z =$ 0 to 0.4, 
 showing the  large  supercluster of galaxies  at  $z \approx 0.3$  is shown in  Figure~\ref{fig:zcone}.

{\em Stripe 82}. \,\, The galaxy redshift   data used in this work is from the central part  of the
 Stripe 82 region of the Sloan Digital Sky Survey  (SDSS-III DR12)
 \citep{Stoughton02,Alam15}.
The data was taken with the multi-fiber spectrograph mounted on the Sloan 2.5 m telescope  
at Apache Point Observatory, NM.
 Stripe 82 is 
 currently  the largest three-dimensional 
 spectroscopic sample of galaxies  with a  high sampling density, and  photometrically up 
 to two magnitudes deeper than the standard SDSS  images. Stripe 82 covers a $\sim270$ degree$^{2}$ 
 area along the celestial  equator in the Southern Galactic Cap and spans $310 \deg < RA <
 59\deg$ and $-1.25\deg \le Dec \le 1.25\deg$.  A major supercluster  
 was noticed  while  exploring  the  large-scale distribution of   
 galaxies surrounding   an unusual filamentary, merging  
 galaxy cluster  ZwCl 2341.1+0000 ($z = 0.27$). This  massive  cluster  also  shows   large-scale diffuse 
  synchrotron  radio  emission  from structure formation shocks, and  thus is  a  prominent target of 
 several studies \citep{2002NewA....7..249B,2009A&A...506.1083V,2013MNRAS.434..772B}.

\begin{figure*}
  \begin{minipage}[b]{.49\linewidth}
    \includegraphics[width=1\linewidth]{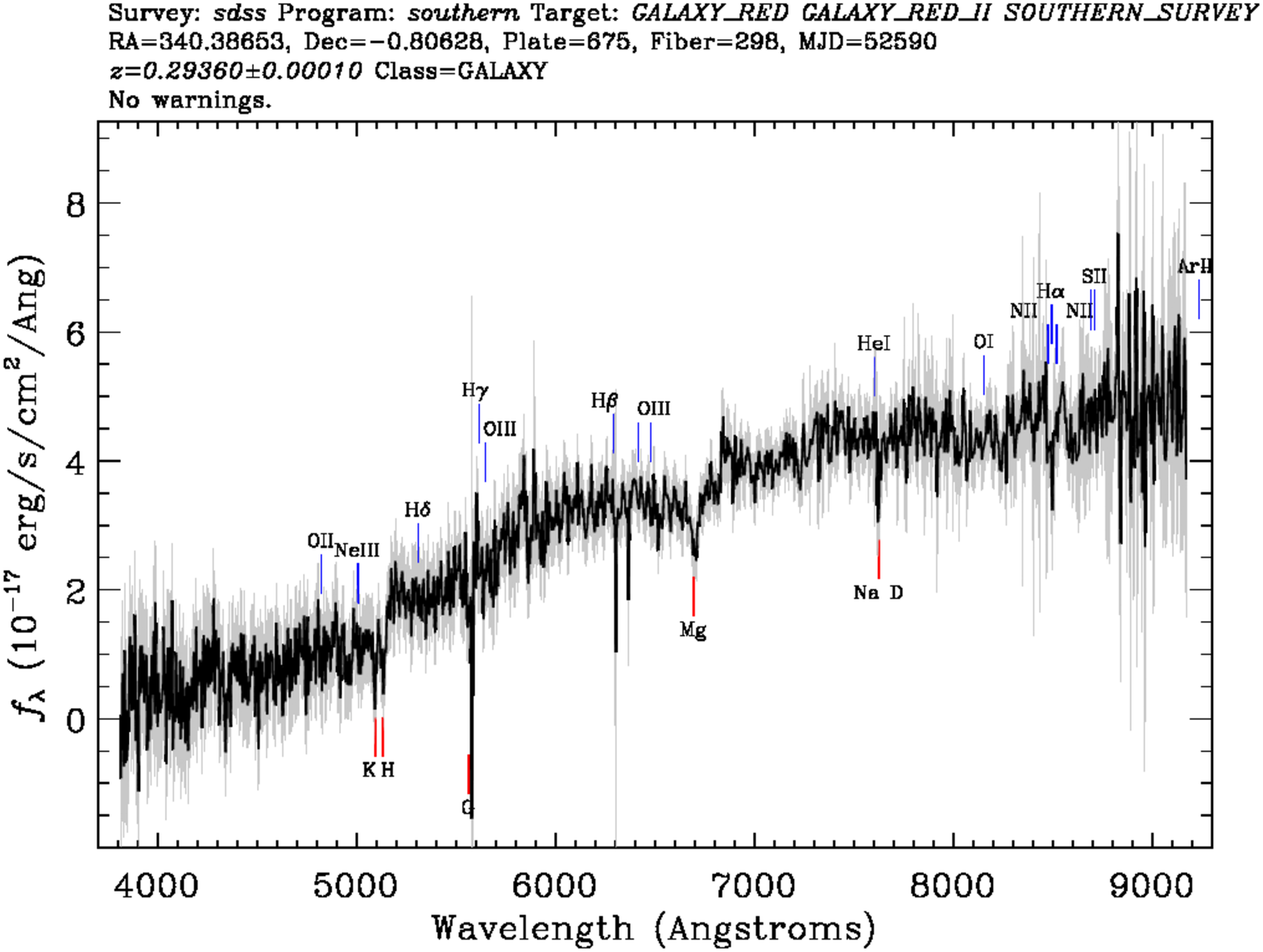} 
  \end{minipage} 
  \begin{minipage}[b]{.49\linewidth}
    \includegraphics[width=1\linewidth]{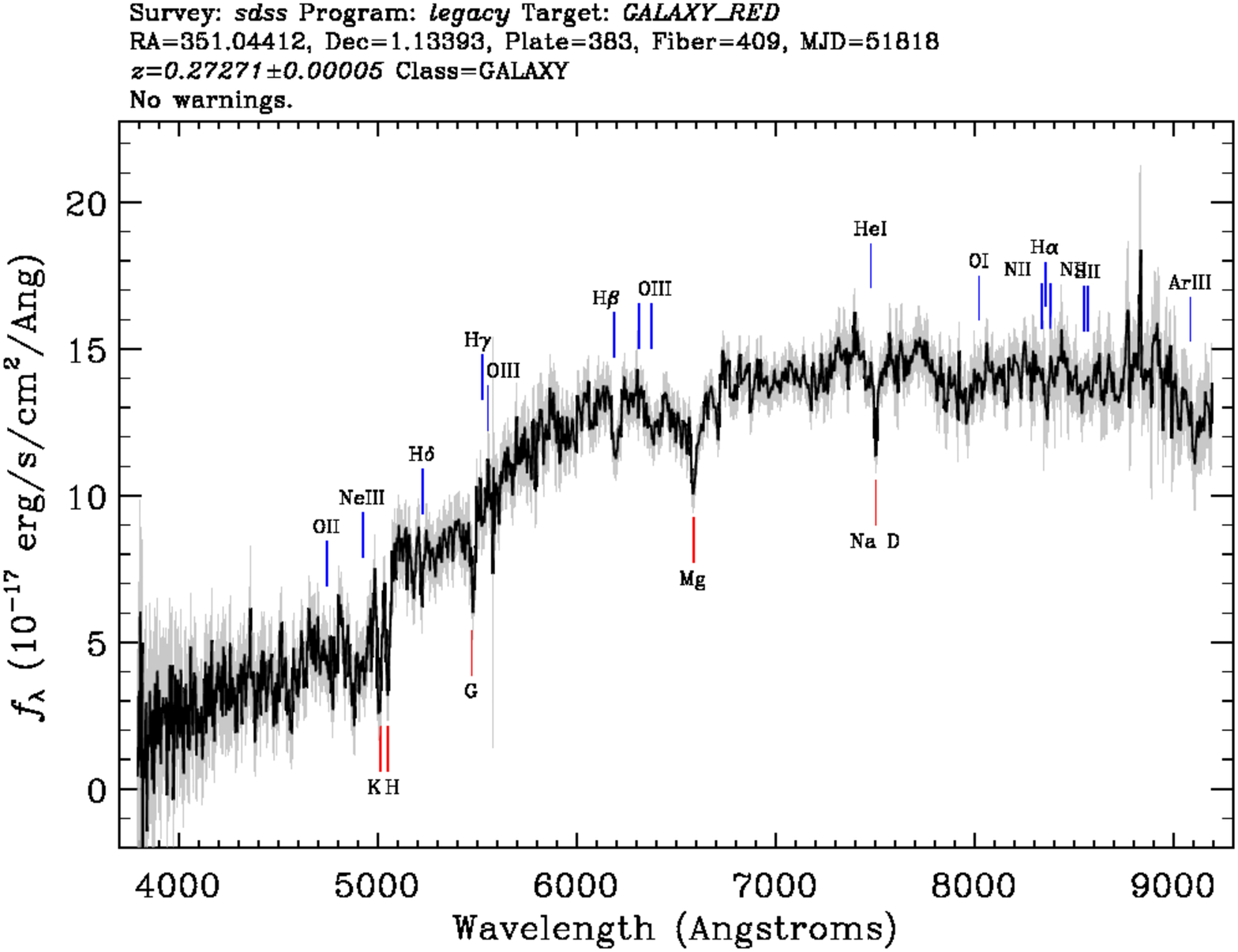} 
  \end{minipage} 
  \begin{minipage}[b]{.49\linewidth}
    \includegraphics[width=1\linewidth]{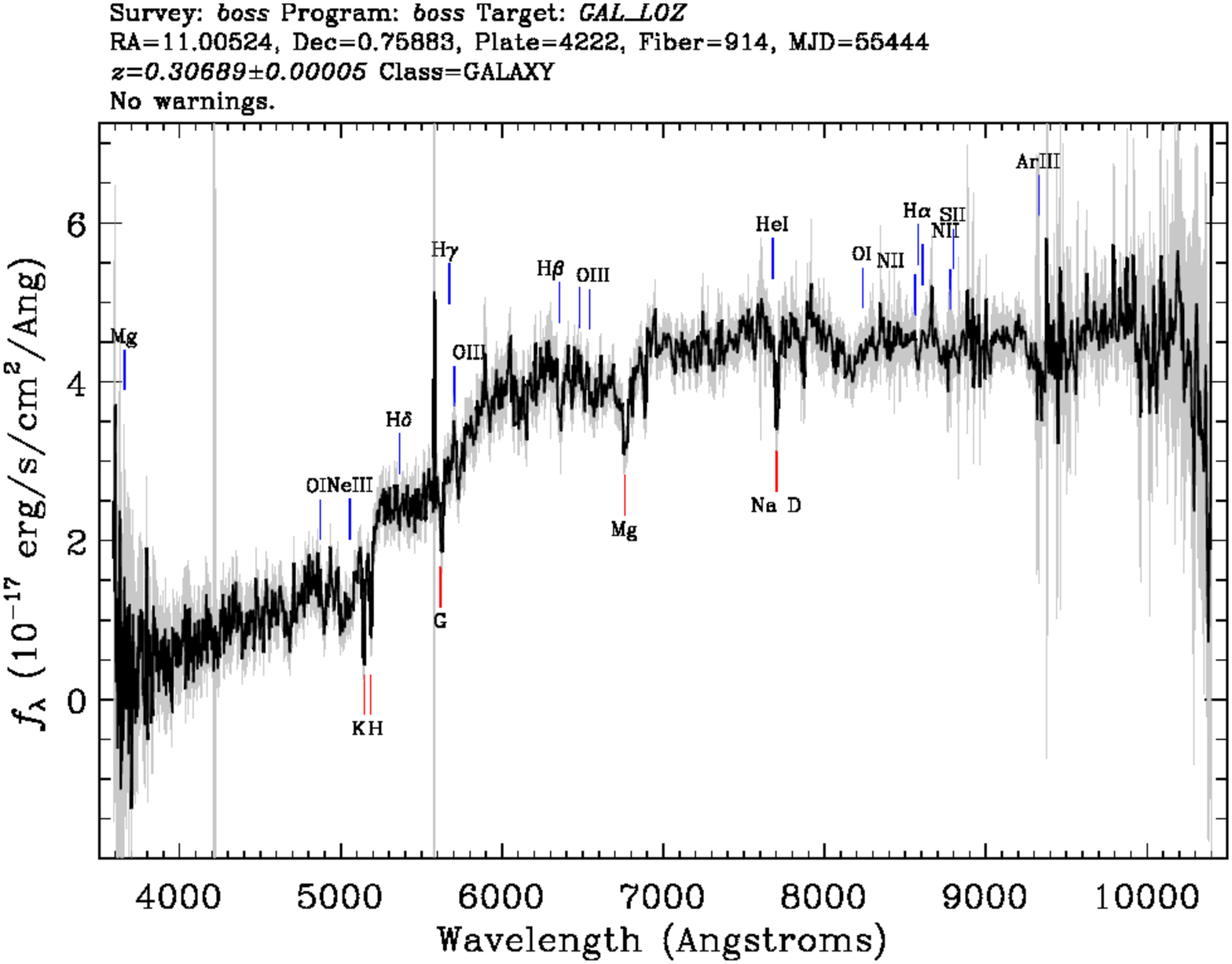} 
  \end{minipage}
  \hfill
  \begin{minipage}[b]{.49\linewidth}
    \includegraphics[width=1\linewidth]{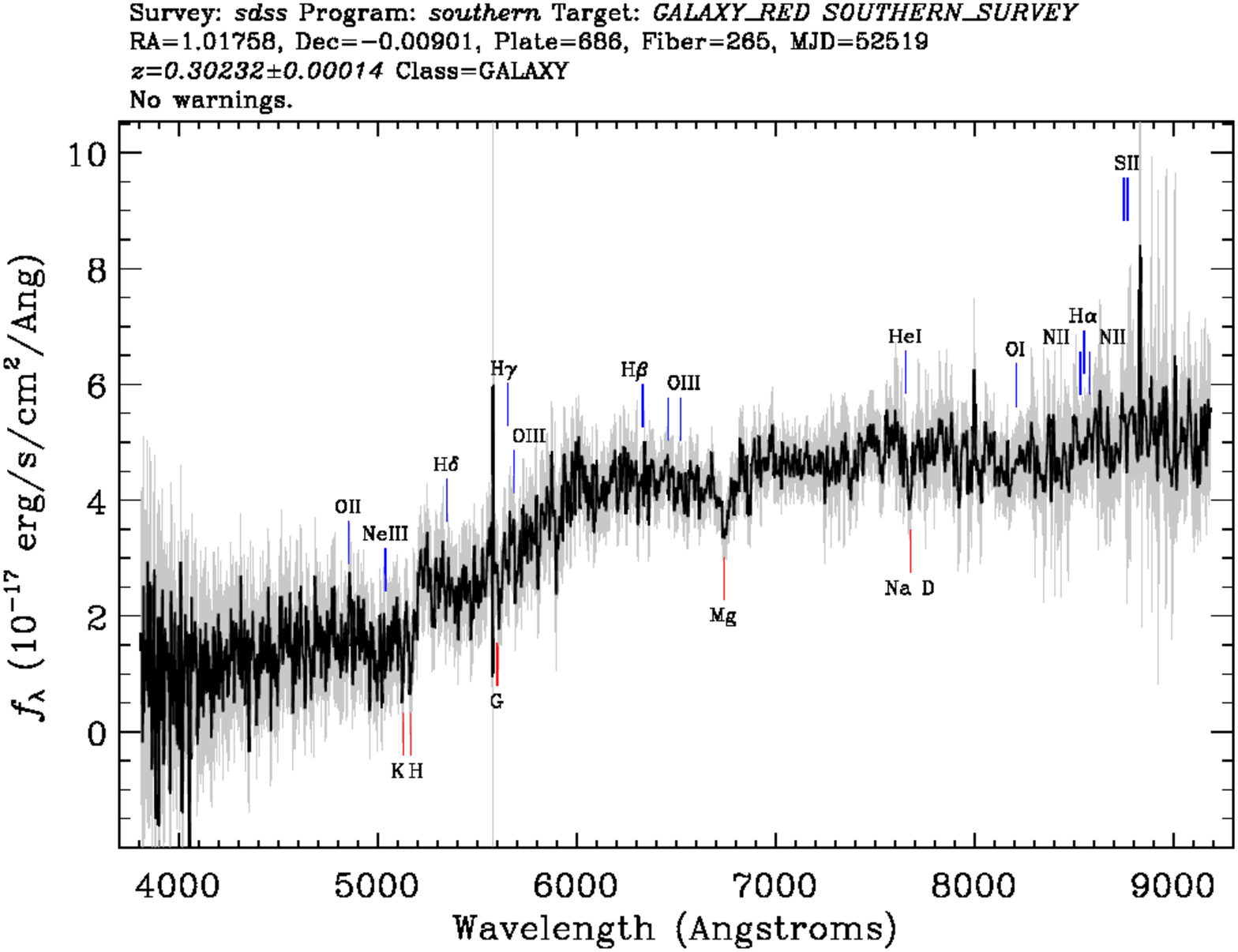} 
  \end{minipage}
  \caption{Spectra of four galaxies from our samples, taken from SDSS.
  The header of each spectra  gives the sky position of the galaxy and  its  redshift with errors. \label{fig:spec}}
\end{figure*}

The entire Stripe 82  is not uniformly sampled (spectroscopically), but contains data from several projects.
We use two largest samples of spectroscopic  galaxies: the  LOWZ sample and the LBS (LEGACY-BOSS-SOUTHERN)
sample described in more details below. We  have 
\textit{k}-corrected  the absolute magnitudes of  galaxies  from both samples by
using the K-CORRECT version 4.3 software \citep{2007AJ....133..734B}. The \textit{k}-corrected 
magnitudes are calculated for  rest frame at $z = 0.33$.  We have retained  only 
those galaxies in our analysis that have
clean photometry, clean spectra, and redshift errors $<$1\%. The spectra of a few galaxies
from our samples are shown in Figure~\ref{fig:spec}.
The apparent  magnitudes   are SDSS {\em cModel} magnitudes corrected for extinction. 
{\em cModel} magnitudes are  composite model magnitudes, calculated from the
best-fitting linear combination of a de Vaucouleurs and an exponential luminosity profile.

\subsection{LOWZ Spectroscopy Data from SDSS DR12} 

\begin{figure*}
\centering
\includegraphics[scale=0.8]{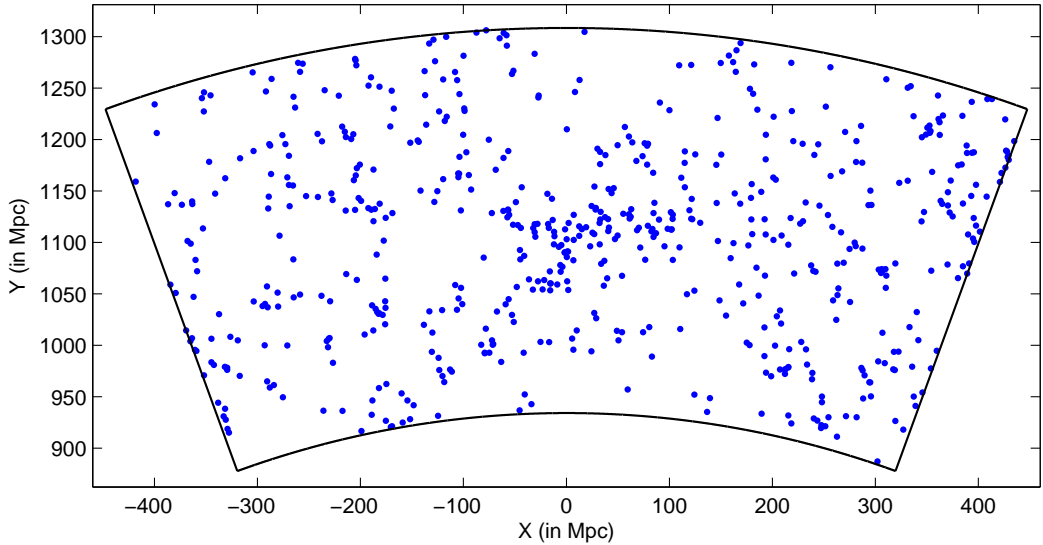}
\caption{Distribution of 625 galaxies in the subsample of LOWZ volume-limited sample in comoving coordinates,
projected on the 2-dimensional X-Y plane. The small depth in the Z-direction (along decl.) has been  suppressed.
\label{fig:vollimdist}}
\end{figure*}

In order to evaluate  the properties of the new supercluster (at a mean $z = 0.288$) and 
explore the  cosmic web  of  galaxies, clusters, and voids  surrounding  it,  
we first extracted  LOWZ  data  for  galaxies from  the Baryon Oscillation Spectroscopic Survey (BOSS)
\citep{Ahn12} of SDSS DR12,   within  a wedge of  R.A.  and decl. range  $336^\circ \le RA \le 16^\circ$,
$-1.25^\circ \le Dec. \le 1.25^\circ$  (limits of  survey) and  
spectroscopic redshift range $0 \le z \le 0.6$, within the Stripe 82  central region.
BOSS targets LOWZ ($z \le 0.6$) galaxies using a set of color-magnitude cuts that follow 
the predicted evolution of a passively evolving stellar population with redshift. The selected 
galaxies are the brightest and reddest of the low-redshift galaxy population, and the targeting 
cuts are similar to those of SDSS-I/II Cut-I Luminous Red Galaxies (LRGs). The LOWZ sample 
extends the SDSS-I/II LRGs by selecting fainter galaxies ($16 < r < 19.6$), thereby 
increasing the number density. 
Due to a bug in the target selection, LOWZ galaxies were incorrectly targeted during the initial
stages of the BOSS survey (affecting all data taken until June 2010). In order to form  a uniformly
targeted sample of LOWZ galaxies, we use  only those galaxies whose TILEID\,$\ge10324$.
A volume-limited galaxy sample was constructed by restricting the
cModel $r-$band absolute magnitude to  $M_r \le -21.53$ and the  redshift to $z \le 0.33$.
From this volume-limited sample we make a subsample with $336^\circ \le RA \le 16^\circ$,
$-1.25^\circ \le Dec. \le +1.25^\circ$ and  $0.23 \le z \le 0.33$,  which is centered on 
the Saraswati supercluster (hereafter called the `analysis wedge/region'). This way we get a total of 
625 spectroscopic galaxies in this subsample. For further analysis we  converted the SDSS angular coordinates ($\alpha, \delta$) 
and redshifts ($z$) of galaxies  to  comoving cartesian
coordinates ($X, Y, Z$). The  two-dimensional distribution of these 
galaxies in comoving  coordinates  is shown in Figure~\ref{fig:vollimdist}.

\begin{figure*}
\centering
\includegraphics[scale=0.8]{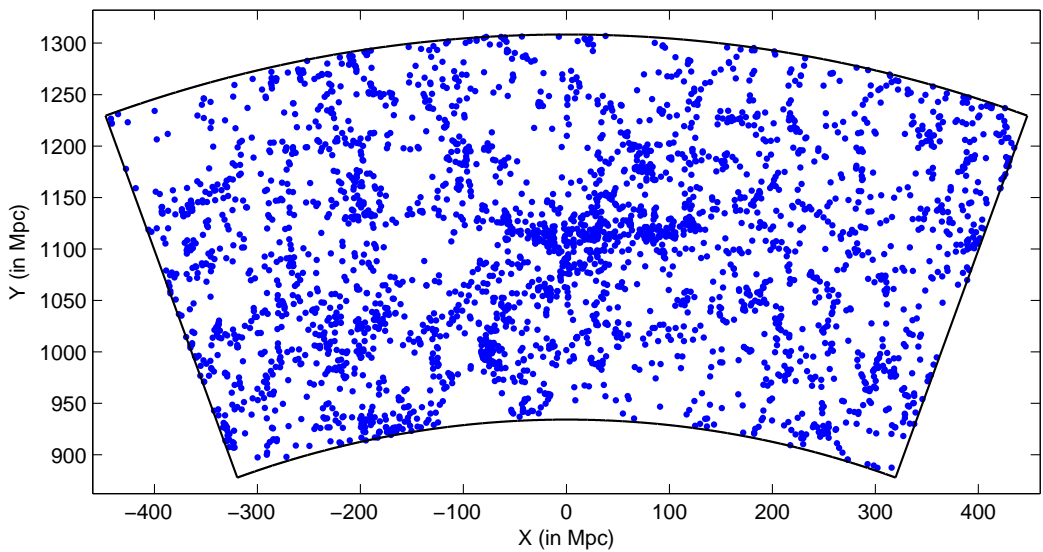}
\caption{LEGACY-BOSS-SOUTHERN (LBS) sample of 3016 galaxies plotted in comoving coordinates, 
projected on the 2-dimensional X-Y plane.  The small depth in the Z-direction (along decl.) 
has been suppressed.
\label{fig:LBS}}
\end{figure*}

\subsection{LEGACY-BOSS-SOUTHERN Data}

The  LOWZ data is uniformly sampled but it is fairly  sparse. In order to
obtain higher density of galaxies within our analysis  wedge (as above),
we combine  all available  spectroscopic galaxies with clean spectra and clean photometry 
taken from ``LEGACY'', ``BOSS,'' and  ``SOUTHERN'' programs of the SDSS-III DR12 database. Only
these  three programs were  selected  because the galaxies in  these programs  were deemed 
to be more or less  uniformly  distributed within our sample region.   
This gives a total of 3016 galaxies, approximately five times the size of the LOWZ sample. 
Hereafter, we shall refer to this  larger sample of 3016 galaxies as the  LBS sample.  
Figure~\ref{fig:LBS} shows the distribution of  galaxies from  LBS sample within our
 wedge in comoving coordinate space.

In both samples, a  prominent,  large-scale ($\sim 200$~Mpc) 
density enhancement of galaxies at   mean $z \sim 0.28$,  
$X \approx  0$ Mpc, and $Y \approx  1100$ Mpc  is clearly visible,  which identifies  the 
Saraswati supercluster,  the focus of our present study. 

\section{Method of Analysis}

Superclusters are the primary large-scale galactic structures  made  of the concentration 
of  smaller galaxy groups and clusters. The identification of  superclusters  
requires  extensive  spectroscopic redshift data taken over vast areas of the sky, 
targeting galaxies in both high- and low-density regions of the cosmic web.
There are generally  two widely used methods to identify the superclusters;
firstly, identifying  overdense structures  based on the Friends-of-Friends (FoF) 
algorithm \citep{Huchra82,Martinez02} with
a particular linking length, which can be applied both to individual galaxies and to groups and 
clusters of galaxies, and secondly, using the  smoothed  density field  
approach \citep{2014dmcw.book.....E}.

We have adopted  the FoF  approach to identify  overdense  structures in the  SDSS  galaxy samples.
In FoF algorithm, a linking length ($l$) is chosen for the galaxy distribution.
If the separation
between two galaxies is less than or equal to $l$, these two galaxies
are considered linked and part of the same group (cluster), otherwise not. Defined in this manner 
the size of a supercluster naturally becomes sensitive to the linking length used to determine it.
The $l$  for a   given point  distribution (galaxies) is generally chosen
such that  maximum number of clusters are obtained given a predetermined limit to
the minimum number of galaxies  required in a cluster.  

\section{Results}
\subsection{Identifying  Clusters and  Superclusters in SDSS Samples}

\subsubsection{Clustering Analysis - LOWZ Galaxies}

\begin{figure*}
\centering
\includegraphics[scale=0.66]{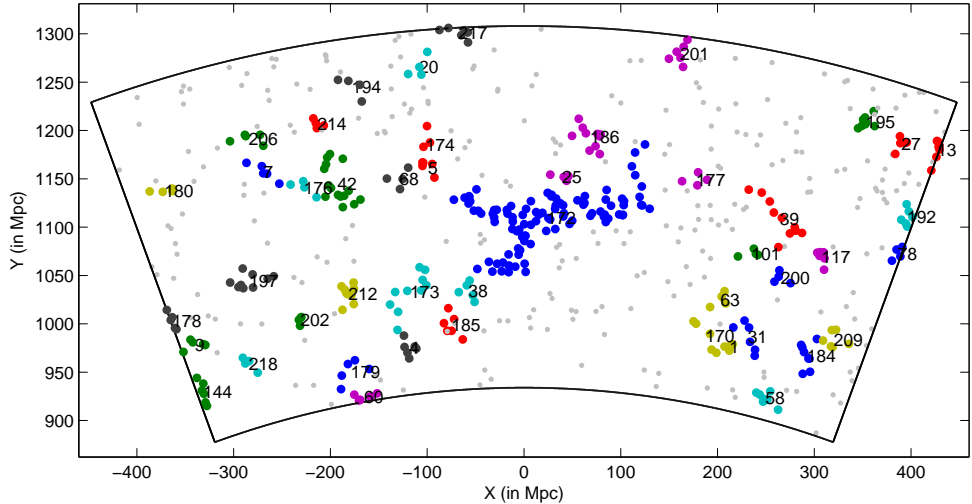}
\caption{Clusters found in the LOWZ subsample by the FoF algorithm for 
comoving linking length of $l = 19$ Mpc. The minimum number of galaxies in a  cluster is four. 
The clusters are shown in colors and  marked with index numbers. In spite of
sparse sampling the largest    cluster  identified  is the Saraswati 
supercluster (index number 172) near the center  shown  in blue,  
containing  $\approx 100$ galaxies. Gray dots are  galaxies that are not part of 
any FoF cluster.
\label{fig:LOWZclust}}
\end{figure*}

We will now perform the cluster analysis (finding clusters) on the subsample of 
625 galaxies from the LOWZ data.  To classify clusters, the FoF algorithm is applied. 
The maximum number of  46  clusters, with  at least four galaxies in a cluster,  
is  obtained at comoving linking length $l \approx 19$ Mpc. Hence we used $19$ Mpc 
as the linking length for the cluster analysis. This  reflects  the sparse sampling of LOWZ data.  
 The largest cluster found this way  is the Saraswati supercluster with  100 galaxies in it. 
 Figure~\ref{fig:LOWZclust} shows all  46  FoF clusters, and  the 
 largest  supercluster Saraswati is shown in blue color at the center of the plot.  
 Saraswati  spans  $\sim200$ Mpc of comoving length across our line of sight. 

\begin{figure*}
\centering
\includegraphics[scale=0.755]{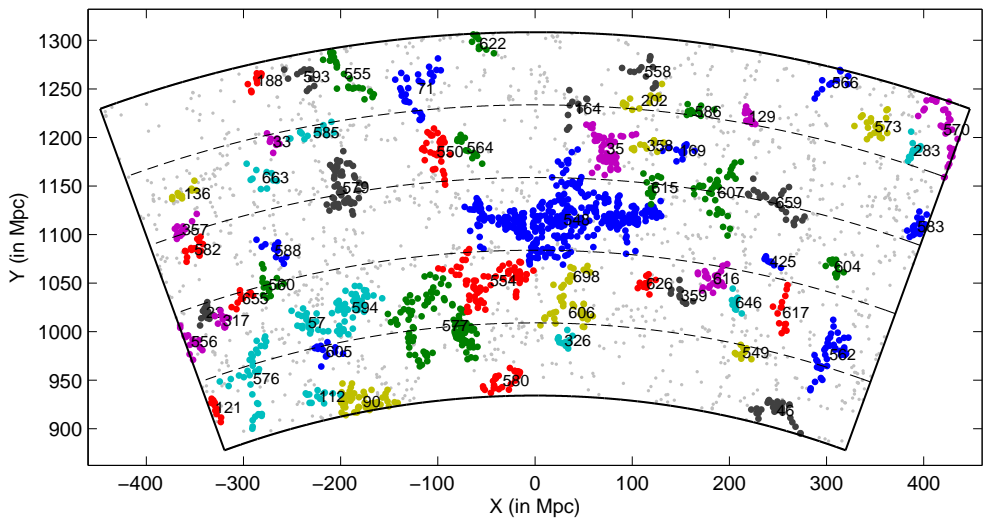}
\caption{Clusters identified in the LBS sample using an FoF algorithm and a comoving linking length $l_{o}=12$ Mpc.
The minimum number of galaxies in each cluster is 10.
Dotted arcs divide the wedge into five redshift bins as mentioned in section 4.1.2.
Different colors and indices indicate different clusters. The largest overdensity of galaxies  found is
Saraswati,   spanning $\sim 200$ Mpc  across (in blue color at the center) 
with $\approx 400$  galaxies in it  and an index  number 548.
Gray dots are galaxies that are not part of any FoF cluster.\label{fig:LBS_clust}}
\end{figure*}

\subsubsection{Clustering Analysis - LBS Galaxies}

\begin{figure*}
\centering
\includegraphics[scale=0.9]{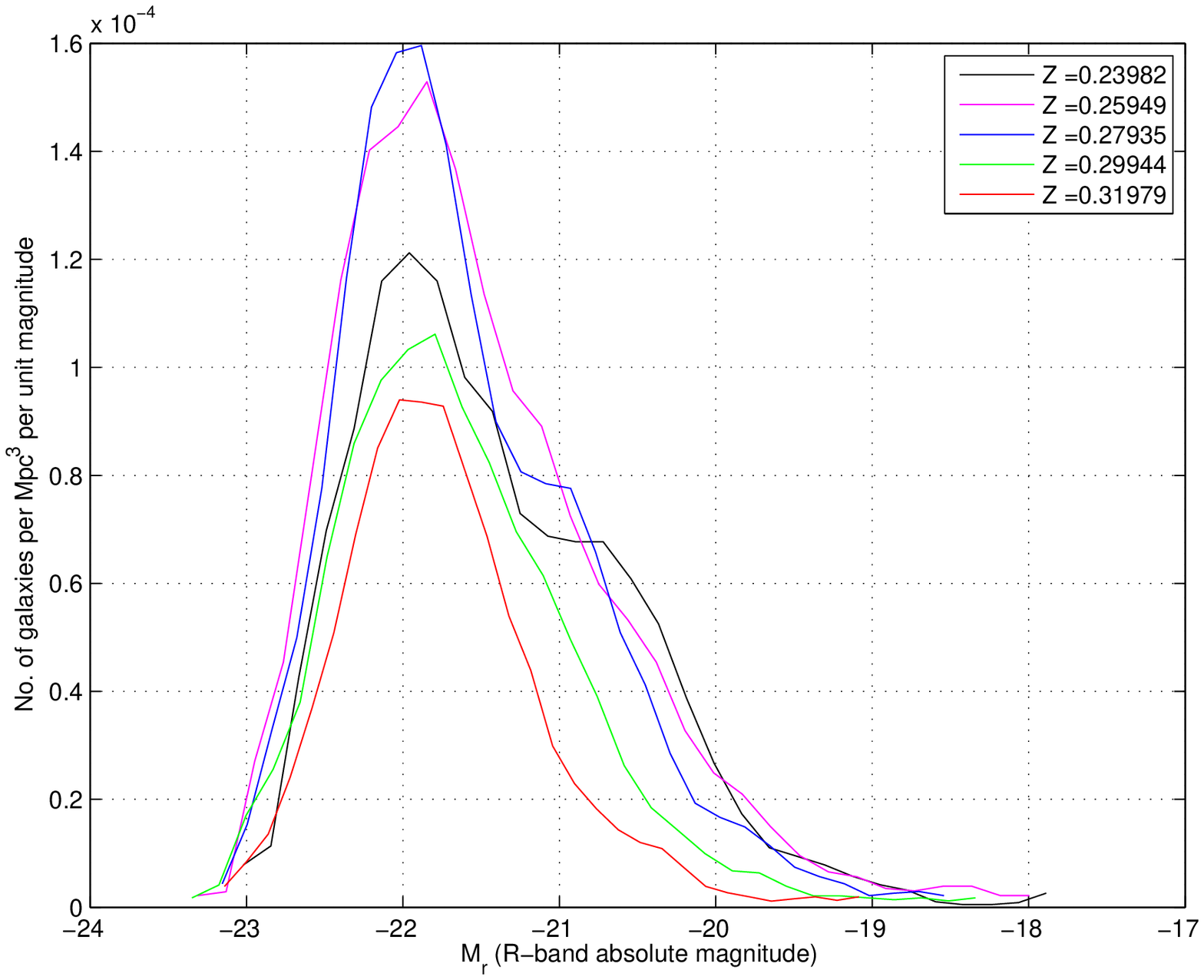}
\caption{Local luminosity functions of LBS sample in 
five different redshift/comoving distance bins. Different color plots  show the
comoving number densities of galaxies per absolute cModel r-band magnitude.
The mean redshift for each luminosity function plot  is shown on top right corner.\label{fig:LF}}
\end{figure*}

Next we  use the FoF algorithm to find  clusters in the LBS sample of 3016 galaxies.
Since this sample is not strictly uniformly sampled in redshift,  we use local
luminosity functions,
i.e. the comoving number density per unit absolute magnitude,
of the galaxies to give proper weights to galaxies so that unbiased clustering
analysis can be performed  using the FoF algorithm. The analysis region  wedge is 
first divided into five constant  comoving  distance bins  and 
the local luminosity functions  of galaxies in each bin are calculated (see Figure~\ref{fig:LF}).
We then assign each galaxy (say $i^{th}$ galaxy),
having an absolute magnitude $M$,
a weight $w_{i}$ according to the values of 
luminosity functions at $M$,

\begin{equation}
 w_{i} = \left [\frac{LF_{1,M}}{LF_{n,M}} \right]_{i}
\label{weight1}
\end{equation}

where $LF_{1,M}$ and $LF_{n,M}$ are the values of local luminosity functions of the lowest
and the $n^{th}$ (the bin in which this $i^{th}$ galaxy is located) comoving distance bins at
magnitude  $M$, respectively. Thus, the weight depends on both the absolute magnitude and the number density of
the  galaxy in the comoving distance bin.
This will assign a weight $w = 1$ to all galaxies in the lowest bin.

Next, we apply  proper weights on the linking length between the two galaxies being linked.
If $l_{o}$ is the selected linking length between galaxy $i$ and galaxy $j$ 
then the weighted linking length $l_{ij}$ will be

\begin{equation}
 l_{ij} = w_{ij} l_{o}
 \end{equation}
 where $w_{ij}$ is the weight between the $i^{th}$ and $j^{th}$ galaxies,
 \begin{equation}
  w_{ij} = \frac{w_{i}^{\frac{1}{3}} + w_{j}^{\frac{1}{3}}}{2}
\label{weight2}
\end{equation}

where $w_{i}$ and $w_{j}$ are the   weights associated with $i^{th}$ and $j^{th}$ galaxies
according to the local luminosity functions (Equation~\ref{weight1} above). 

Using this method, we get the largest number of clusters  at $l_{o} = 12$~Mpc with a minimum number 
of galaxies in a cluster to be 10. Using these parameters, we identify a total of 60  FoF 
galaxy clusters within the LBS analysis region. Once again, the  largest  cluster  
with a maximum number of  linked galaxies, naturally comes out to be the 
Saraswati supercluster with $\sim400$ galaxies in it. 
In Figure~\ref{fig:LBS_clust}, all our identified  clusters are shown with various  
colors and  using different index
numbers.  The Saraswati  supercluster spanning $\approx 200$~Mpc across the 
line of sight ($\approx 9 \deg$) stands out prominently  in the middle of  the plot. Moreover,  
even in  limited  sampling  one  gets  a  clear  visual impression of the  rich tapestry of 
 filaments, galaxy  clusters, and  large voids of $\approx 40 - 170$~Mpc diameter surrounding
the supercluster.

\begin{figure*}
\centering
\includegraphics[scale=0.755]{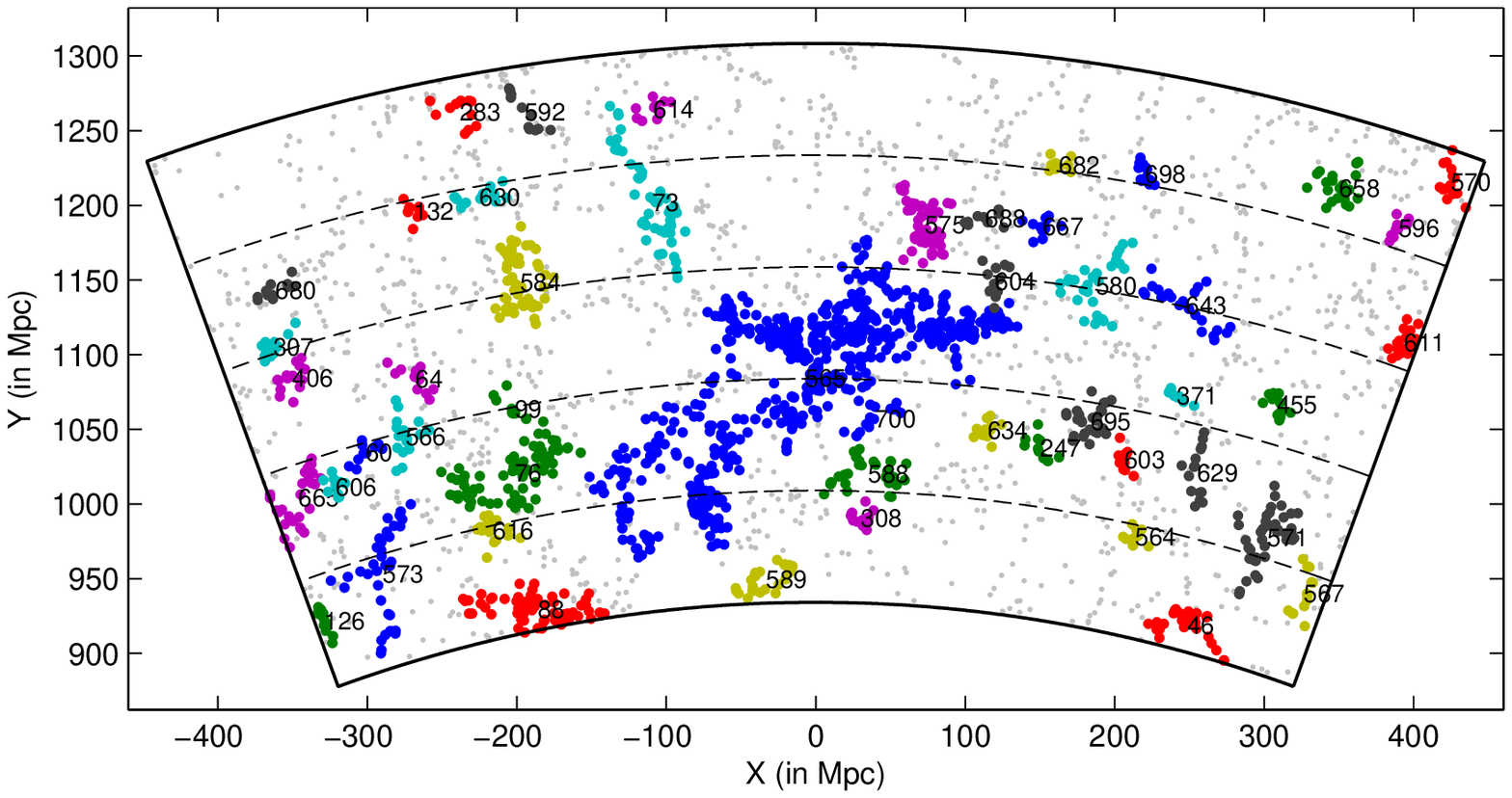}
\caption{
Clusters in LBS sample using FoF algorithm and a fixed (unweighted) linking length of 12 Mpc.
The minimum number of galaxies in each cluster is 10. Different colors and indices indicate different
clusters. This  analysis  extends  the  Saraswati supercluster further toward lower redshifts  by linking  it  with two 
nearby  large galaxy filaments (in blue color at center).
\label{fig:LBS_clust_noweight}}
\end{figure*}

Figure~\ref{fig:LBS_clust_noweight} shows the FoF clusters using a fixed linking length of 12 Mpc and with all weights set equal to 1.
The figure indicates  that unweighted  linking length  further  increases the   extent  of   Saraswati  downwards  by 
linking  it  with  two nearby large galaxy filaments. Whereas  our clustering analysis  with adaptive weighting scheme detects them as
separate structures.

Therefore, based on  this  study,  and   with   the firm  identification of several   
massive   galaxy  clusters    and  voids near   the Saraswati 
supercluster in  sections below, we have very little scope for doubt left that it is a real physical 
structure formed via  cosmological processes, and not  an artifact of 
chance alignment and  projection  effects.

\subsection{The  Affect  of  Redshift Errors and Peculiar Velocities on  Clustering Analysis}

As mentioned in section 2, the spectroscopic redshift errors are less than 1$\%$ for all galaxies in our samples.
The  small  redshift  errors have a negligible affect  on our clustering  analysis. Another important factor is the
Redshift Space Distortion (RSD), which is more significant for galaxies within the virial radius $r_{200}$ of a cluster.  To take into account the
affect of  RSD,  we calculated, for the  LBS sample, using  the virial masses $M_{200}$ and positions of WHL clusters (see Sections 5 and section 6.1), 
the RSD induced by  these clusters in  radial direction on the  suurounding  galaxies  up to  distances  $r_{200}$.
We corrected the comoving distances for all those galaxies which are within angular virial radius of the cluster and within a distance of
($v_{200}/H_o$) from BCG (brightest cluster galaxy) in radial direction using the following equation,

\begin{equation}
  d_{gal} = d_{BCG} + (d_{gal}^{*} - d_{BCG})\frac{r_{200}}{v_{200}/H_{o}}
\label{pec1}
\end{equation}

where $d_{gal}$ is the new comoving distance of galaxy, $d_{gal}^*$ is the old comoving distance of galaxy, $d_{BCG}$ is the comoving
distance of BCG, $r_{200}$ is the virial radius of the cluster, $v_{200}$ is the radial
peculiar  velocity component at $r_{200}$ due to  mass  $M_{200}$ ($v_{200} = \sqrt{GM_{200}/3r_{200}}$),  and $H_{o}$ is the Hubble parameter
at $z =$ 0.

Equation~\ref{pec1} is similar to Equation (1) of \cite{Liivamagi12}. Since our spectroscopic sample is not dense enough to give a high
number density of member galaxies in a cluster, we found only  224 galaxies for  RSD corrections.
The mean, median, and standard deviation of correction $\bigtriangleup d$ to comoving distances of these 224 galaxies are 1.29 Mpc,
0.27 Mpc, and 1.73 Mpc. After applying  these corrections, we
again performed the clustering analysis with weights and found that the major results are  not changed significantly. We again obtain  60  FoF 
clusters in all,  with Saraswati being the largest cluster having 386 galaxies in it,  exactly  the  same as before. 
Reanalysis of the RSD-corrected LOWZ sample also concludes that RSD has negligible effect on  our  clustering analysis.

This shows that the detection of rhe Saraswati supercluster and other clusters is  robust to redshift errors and peculiar motions in our 
data.
\bigskip
\bigskip
\section{Distribution of Known Clusters and  Voids  Near the Supercluster}

In the $\Lambda$CDM  paradigm, clusters of galaxies trace the local extrema in the underlying
distribution of matter -- both dark and luminous, and
since Saraswati is a particularly vast and   overdense structure of galaxies, some 
  clusters or groups of  galaxies should  be found at
 the highest  density peaks. Moreover, in cosmic web  overdense regions are surrounded by almost
 empty voids.   For  quantifying  more objectively  the galaxy distribution  in   overdense
(clusters/groups, filaments) and  the  underdense (voids) regions of the  cosmic web 
around the Saraswati  supercluster,  we cross-match the galaxy distribution  
 with the published  catalogs of  clusters and voids derived  from SDSS data.

We compare the distribution of galaxies in our sample region  with the
 clusters listed in SDSS-III cluster catalog of 
\cite{2012ApJS..199...34W} (hereafter the WHL catalog). In the WHL catalog,
the photometric redshifts of  the BCG of each
cluster are given. The spectroscopic redshifts are also listed but only 
if  spectra  are available in  SDSS. We have used
all available spectroscopic  redshifts  in SDSS-III  for the
 BCGs of WHL  galaxy clusters. For   other clusters (small in number) 
whose BCG spectra are not available, we used   photometric  redshifts. 
In this way,  we identified  238 WHL clusters  in all  within our  analysis region.
The WHL cluster catalog is $\approx75\%$ complete for clusters of
$M_{200} > 0.6 \times 10^{14} M_{\odot}$
and redshifts $z < 0.42$, and $\approx100\%$  complete for clusters with
$M_{200} > 2 \times 10^{14} M_{\odot}$ and $z < 0.5$ \citep{2012ApJS..199...34W}. 

\begin{figure*}
\centering
\includegraphics[scale=0.8]{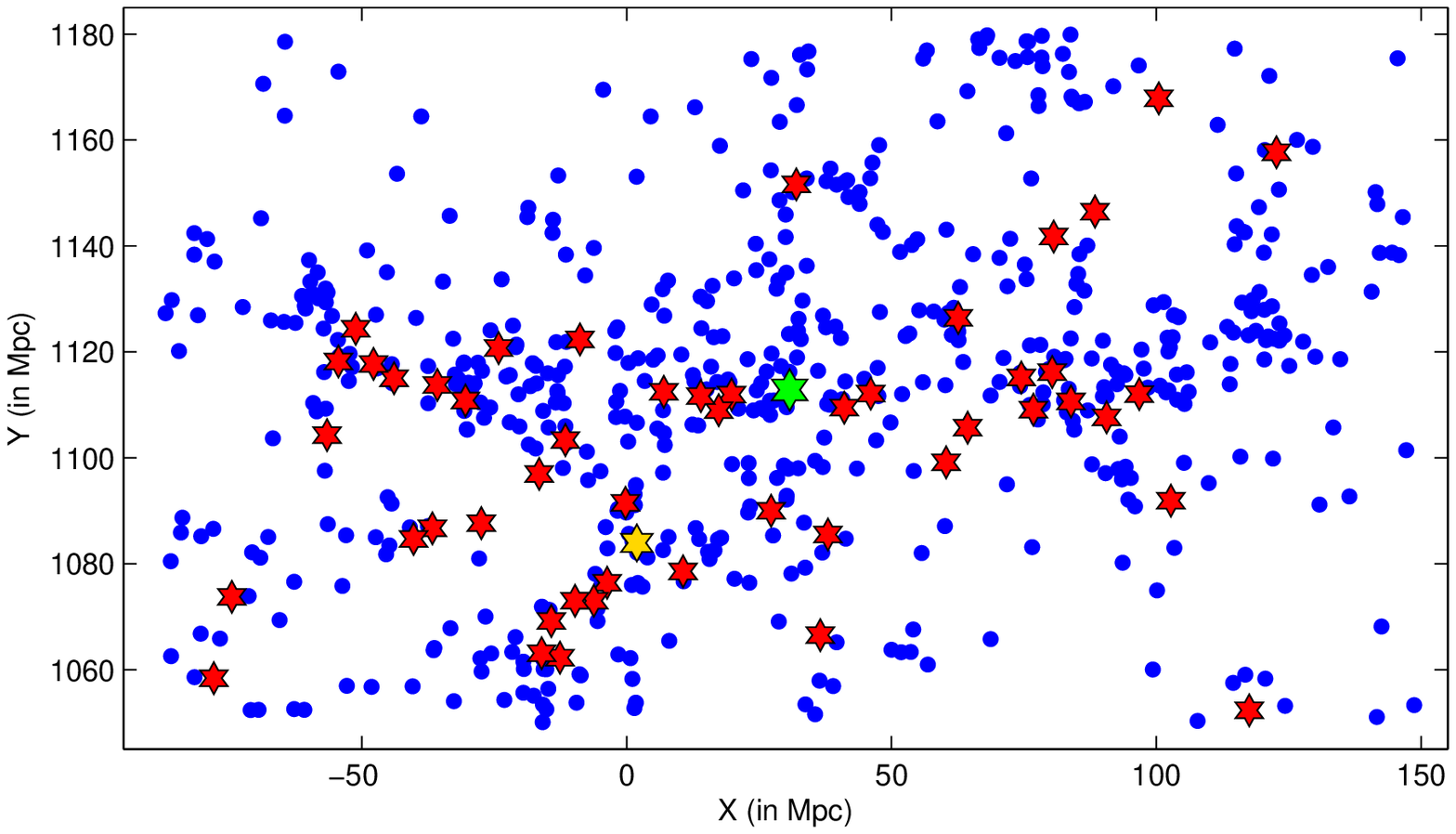}
\caption{In this figure  we show galaxy  clusters near the Saraswati supercluster. All cataloged  
WHL  Clusters (red stars) are  plotted over the galaxies from the LBS sample (blue dots). The location  of the 
most massive cluster  Abell 2631  is  shown with a green star  and the second-most 
massive cluster ZwCL 2341.1+0000  by a yellow star (see Figure~\ref{fig:optical_decam} and 
Table~\ref{table1}).  \label{fig:LBS_WHL}}
\end{figure*}

Out of  the 238 WHL clusters within  our analysis region  there is a 
  major  concentration  of  48  clusters 
 within  90 Mpc comoving distance from the center of Saraswati,  and    43  of these are 
 associated with the  filament/wall-like  main structure of the Saraswati  
  supercluster (Figure~\ref{fig:LBS_WHL} and \ref{fig:saras}).   
  The data  on   these  clusters are  given   in Table~\ref{table1},  listed  in decreasing
  order of  their mass  ($M_{200}$). 
The first cluster in the  list is Abell 2631 ($z=0.277$) and the second one  
is  ZwCl 2341.1+0000 ($z=0.269$), both well-known clusters.   We have further analyzed the 
properties of  these  WHL  clusters below and  used them to estimate the
total mass  and  overdensity  of  Saraswati.




For the study of voids, we use  the LOWZ void  
catalog of \cite{2016MNRAS.461..358N} based on the BOSS data of SDSS
 Data Release 11 (hereafter `Nadathur  voids').
Nadathur voids are identified using the {\em Voronoi Tessellation} and {\em Watershed} 
algorithms. These are disjointed voids -– independent underdense regions of 
space that do not overlap with each other. We have identified a total 
of  24   voids within our analysis  region whose
comoving radii range from  $\approx 20$ Mpc to $\approx 86$ Mpc.

\begin{figure*}
\centering
\includegraphics[scale=0.8]{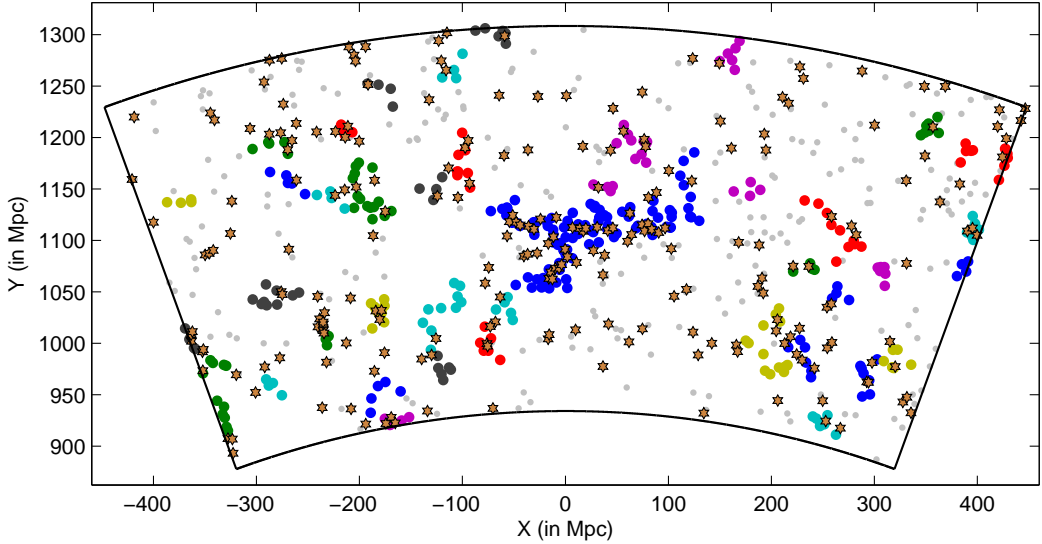}
\caption{Here we plot the  WHL clusters (denoted by stars) along with  FoF clusters  
identified with LOWZ galaxy data. The high density of  clusters near  the Sarasawati 
supercluster is  very clear.\label{fig:LOWZclustWHL}}
\end{figure*}

\begin{figure*}
\centering
\includegraphics[scale=0.8]{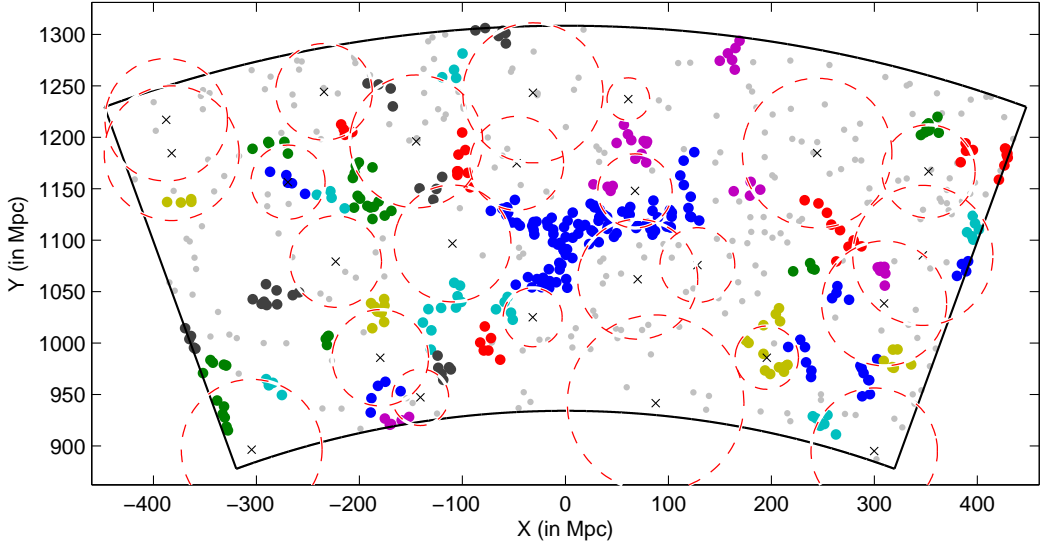}
\caption{Voids distribution in LOWZ subsample. The  FoF clusters are 
shown in various colors.  
Red dashed circles are voids in projection, black crosses are
void centers,  and the radii of circles are equal to the  effective radii of voids as given in
the Nadathur void catalog. \label{fig:LOWZclustVoid}}
\end{figure*}

\textit{Clusters and Voids--LOWZ Sample} \,:\,
the distribution of WHL clusters  in and around 
Saraswati  is shown in Figure~\ref{fig:LOWZclustWHL}
where clusters are plotted  with star symbols. This  figure shows  
that  number  density  of  galaxy  clusters within the Sarasawati region is 
much  higher  compared to other lower density regions. A similar result is 
obtained using the LBS galaxy
sample discussed below.
Figure~\ref{fig:LOWZclustVoid} shows the distribution of voids within our analysis  region.
The circles (red dashed) show the  voids, cross markers show the centers  of voids and the
radii of voids  are the  effective  radii
as given in the Nadathur  void catalog. We can easily see that many  voids 
surround  the Sarasawati supercluster,  which is  expected because 
superclusters are  always  surrounded by  voids.

\begin{figure*}
\centering
\includegraphics[scale=0.8]{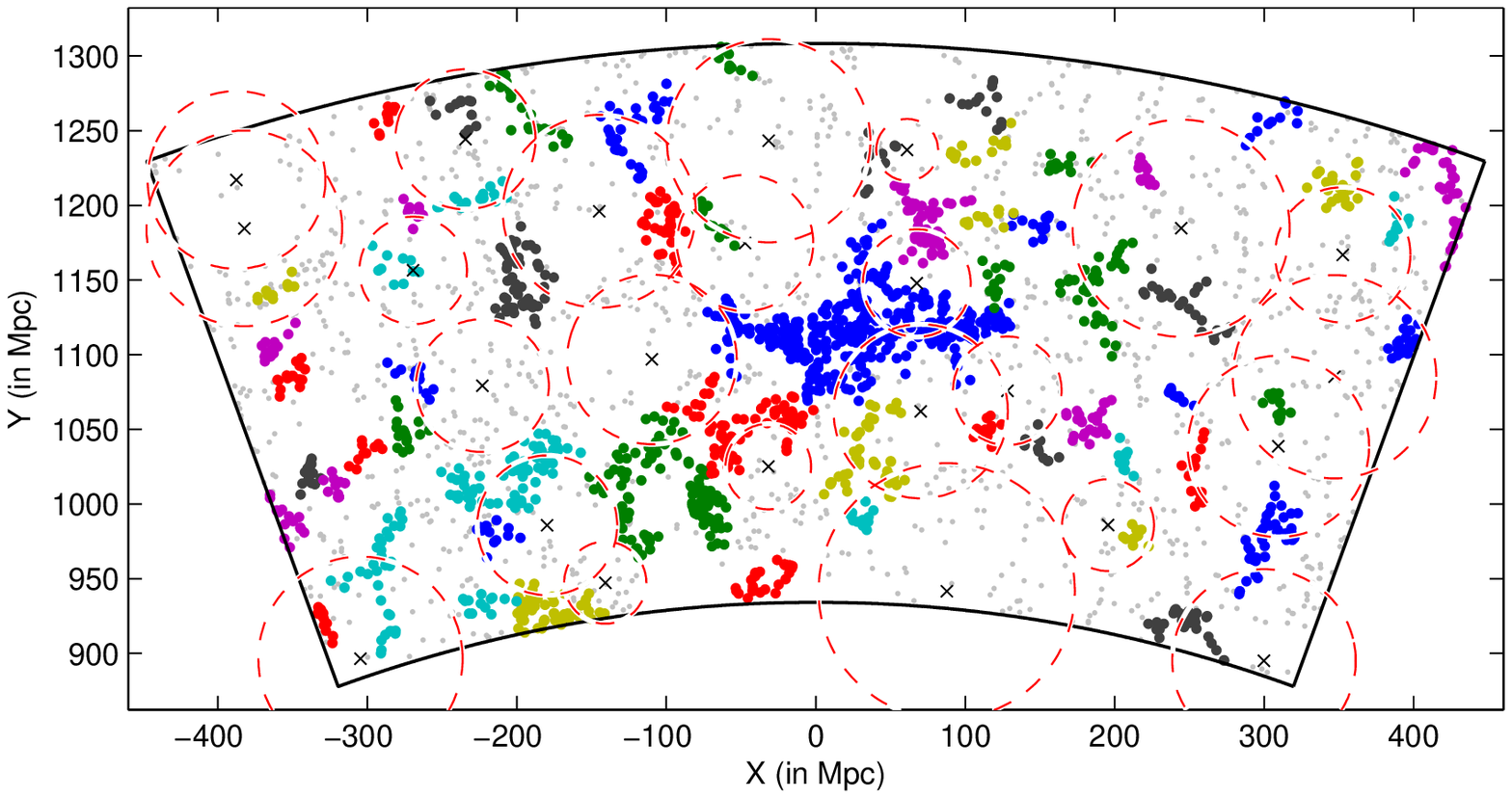}
\caption{In this figure we show   voids  in the Saraswati supercluster region. Cataloged 
voids \citep{2016MNRAS.461..358N} are plotted over the LBS galaxy sample. The FoF clusters are shown in various colors. 
Red dashed circles are voids in projection, black crosses are
void centers,  and the radii of circles are equal to the effective radii of voids as given in
the Nadathur void catalog.\label{fig:LBS_void}}
\end{figure*}

\textit{Clusters and Voids--LBS Sample} \,:\,
similarly,  the  WHL clusters and  Nadathur voids are  plotted on  the  LBS galaxy sample, shown 
in  Figure~\ref{fig:LBS_WHL} and Figure~\ref{fig:LBS_void}. 

The high density of galaxies within Saraswati in both  galaxy  samples (LOWZ and LBS),
high density of  WHL clusters within Saraswati,   and the  presence of  
large voids surrounding  it  strongly  support the physical existence of the
Saraswati supercluster.

\section{Properties of the Saraswati Supercluster}

\begin{figure*}
\centering
\includegraphics[scale=0.265]{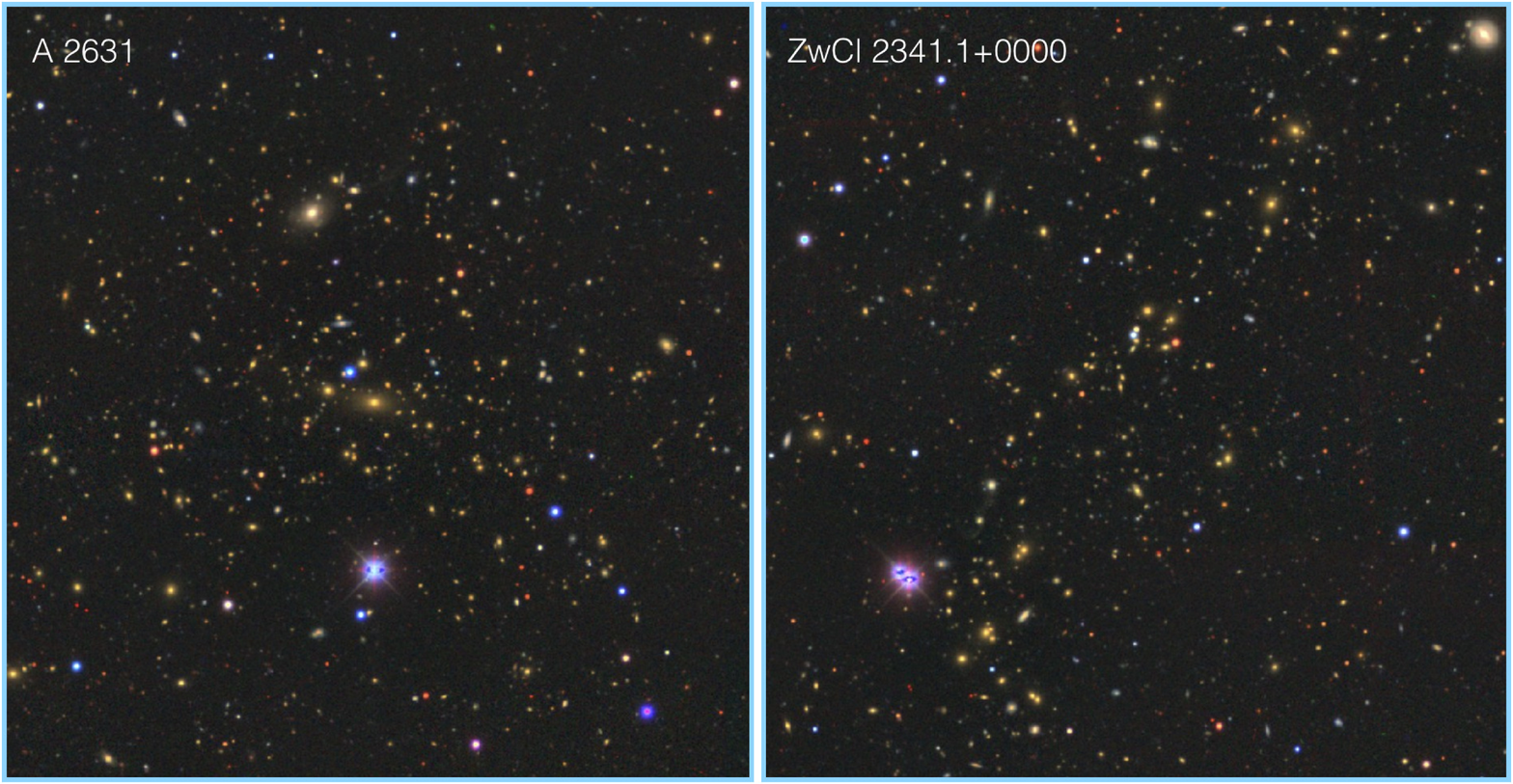}\\
\caption{Optical images of the two  most massive clusters  of the Saraswati supercluster; 
with the relaxed-type  Abell 2631 ($z=0.277$) on left and the highly filamentary  ZWCl 2341.1+0000  ($z=0.269$) on the right, 
taken  from the DECam legacy survey (DECaLS).  Each  image is  $\approx10^{'}\times 10^{'}$ across, 
corresponding roughly to 2.5 Mpc on a side.}
\label{fig:optical_decam}
\end{figure*}

\subsection{Mass Distribution}

We have identified the Saraswati  supercluster as a high density region of galaxies 
at mean redshift $z \sim 0.28$. A  search through the Abell cluster catalog \citep{ACO89} 
shows that  it is 
centered on the extremely rich (Abell richness class $R = 3$),  
massive ($M_{500} \approx  10^{15} M_{\odot} $) and hot
($T_{e} = 8$~KeV) galaxy cluster Abell 2631 at redshift $z = 0.277$. 
Abell 2631 is  detected  strongly in X-rays and also shows a  significant 
Sunyaev--Zel'dovich decrement signal on the CMBR  \citep{2012ApJ...75}.
We show below that the dominant Abell 2631 and  a few more  nearby massive clusters most likely  form
the dense, bound core of the Saraswati supercluster.
Another unusual galaxy cluster ZwCl 2341.1+0000 ($z = 0.269$), a  highly filamentary,  
multiply merging  system, is located $\sim 45$ Mpc to the south in a
filamentary spur joining the main Saraswati concentration. This one is 
the second-most massive  cluster in  the supercluster region.
The  faint,   non-thermal  synchrotron radio emission  on a Mpc scale discovered in the ZwCl 2341.1+0000 
cluster \citep{2002NewA....7..249B,2009A&A...506.1083V,Giovannini10}
is a clear signpost of the infall and merger dynamics of several galaxy
groups during the first phase of the cluster formation process \citep{2013MNRAS.434..772B}. 
The optical images  of  clusters Abell 2631 and ZwCl 2341.1+0000 are shown in 
Figure~\ref{fig:optical_decam}.

The main body of  Saraswati supercluster contains at least   43 massive
clusters/groups   of galaxies identified from the WHL catalog.  
The total  mass of a supercluster  is  indicated by the  number of 
massive clusters it contains \citep{2014A&A...567A.144C}.
The total bound halo mass of these 43 clusters is estimated to 
be $\approx 2\times10^{16} M_\odot$, as explained below.
From the known masses of clusters, through X-ray and weak-lensing
scaling relations, and  their  richness, \cite{2012ApJS..199...34W}
found a correlation between the virial mass $M_{200}$ and the cluster richness parameter  $R_{L\ast}$ of
their  WHL  clusters, given by
\begin{equation}
 log \, M_{200} = (-1.49 \pm 0.05) + (1.17 \pm 0.03) \, log \, R_{L\ast},
 \end{equation}
 where $M_{200}$ is in units of $10^{14} M_{\odot}$.
 
 Using  this relation, we  estimate the mass of each cluster
 within  $r_{200}$  (see Table~\ref{table1}).  However, a significant amount of 
 matter must lie beyond the
 virial radius of a cluster,  which needs to be accounted for. Therefore,
 this  estimate of virial mass  $M_{200}$  is further scaled up  to calculate the 
 bound halo mass  $M_{halo} \approx M_{5.6}$ of each  cluster;  where 
 $M_{5.6}$ denotes  the mass inside  $r_{5.6}$, the radius within which the mass overdensity is
 5.6 times the critical density of the  universe.  Several  studies find  
 that the  bound halo mass  of a cluster is  close to
 $M_{5.6} \sim 2.2 \times M_{200}$ \citep{2005MNRAS.363L..11B,Rines06,2013ApJ...767...15R}. 
 This way   we estimate the total  halo  mass  of  43  main  clusters  to 
 be  $\sim 2\times10^{16} M_\odot$.  We point out that this is strictly a lower limit as the 
 full extent of  the Saraswati   supercluster is    not  known  yet 
 (due to the decl. limit of  SDSS on Stripe 82), 
 and some fraction of total  mass is still  unaccounted for, which is not  
 bound to the  virializing dark matter  halos.

\begin{deluxetable*}{c c c c c c c c c}
\tabletypesize{\scriptsize}
\setlength{\tabcolsep}{12.5pt}
\tablecaption{WHL Clusters in and around the Saraswati Supercluster\label{table1}}
\tablehead{
\colhead{SrNo} & \colhead{R.A.} & \colhead{Decl.} & \colhead{Redshift} & \colhead{\textit{r}mag} & \colhead{$R_{200}$} & \colhead{$N_{200}$} & \colhead{$M_{200}$} & \colhead{Comv. Dist.} \\
\colhead{} & \colhead{(deg)} & \colhead{(deg)} & \colhead{} & \colhead{} & \colhead{(Mpc)} & \colhead{} & \colhead{(10$^{14} M_{\odot}$)} & \colhead{(Mpc)}
}
\startdata
 \dag{1} & {354.41553} & {0.27137} & {0.2772} & {17.23} & {1.93} & {103} & {10.4544} & {0.000} \\
 2 & 355.89862 & 0.33093 & 0.2694 & 17.93 & 1.62 &  79 & 7.4352 & 40.838 \\
 3 & 356.59955 & 0.74942 & 0.2746 & 17.42 & 1.60 &  51 & 4.6789 & 44.331 \\
 \dag4 & 355.27872 & 0.30925 & 0.2768 & 17.21 & 1.36 &  32 & 2.9718 & 16.843 \\
 5 & 354.40656 & -0.67781 & 0.2876 & 17.51 & 1.39 &  31 & 2.5886 & 43.159 \\
 6 & 356.86499 & -0.15381 & 0.2639 & 17.91 & 1.33 &  37 & 2.4723 & 68.773 \\
 7 & 350.62363 & -0.37294 & 0.2728 & 18.33 & 1.32 &  34 & 2.4673 & 75.977 \\
 8 & 351.68036 & 1.13423 & 0.2774 & 18.28 & 1.37 &  30 & 2.4354 & 55.740 \\
 9 & 351.88278 & 0.94281 & 0.2788 & 17.14 & 1.27 &  28 & 2.3992 & 51.391 \\
 10 & 356.51950 & -0.18573 & 0.2665 & 17.62 & 1.28 &  27 & 2.1499 & 57.477 \\
 11 & 357.56891 & 0.88373 & 0.2767 & 17.79 & 1.28 &  28 & 2.1138 & 62.379 \\
 12 & 352.86020 & 0.61525 & 0.2739 & 17.84 & 1.20 &  26 & 1.7984 & 33.171 \\
 13 & 357.83884 & 0.61691 & 0.2775 & 17.67 & 1.19 &  23 & 1.6145 & 66.884 \\
 14 & 358.78690 & 0.75341 & 0.2789 & 17.87 & 1.07 &  24 & 1.4692 & 85.906 \\
 15 & 354.03653 & -1.18347 & 0.2650 & 17.75 & 1.10 &  26 & 1.4580 & 54.053 \\
 16 & 352.03851 & 0.18593 & 0.2768 & 17.38 & 1.12 &  16 & 1.3240 & 46.205 \\
 17 & 356.45068 & -1.12771 & 0.2797 & 17.67 & 1.08 &  19 & 1.2245 & 49.083 \\
 18 & 356.19455 & -0.08880 & 0.2674 & 17.98 & 1.05 &  17 & 1.1919 & 50.579 \\
 19 & 354.56683 & 0.11680 & 0.2711 & 17.60 & 1.07 &  24 & 1.1810 & 23.271 \\
 20 & 358.25555 & -0.23089 & 0.2779 & 17.86 & 1.01 &  10 & 1.0581 & 75.367 \\
 21 & 356.01273 & 0.22644 & 0.2714 & 18.34 & 1.00 &  18 & 1.0368 & 37.667 \\
 22 & 356.86264 & 0.52773 & 0.2729 & 17.74 & 1.00 &  12 & 1.0255 & 50.122 \\
 23 & 353.99393 & -0.48434 & 0.2700 & 17.79 & 0.99 &  15 & 1.0180 & 31.727 \\
 24 & 356.32980 & -0.05331 & 0.2665 & 18.29 & 0.98 &  19 & 1.0080 & 54.668 \\
 \dag25 & 355.10202 & -0.09299 & 0.2761 & 18.09 & 0.98 &  14 & 0.9881 & 15.627 \\
 26 & 358.93271 & -0.00393 & 0.2752 & 17.95 & 0.98 &  16 & 0.9868 & 87.932 \\
 27 & 353.62631 & -0.98501 & 0.2772 & 17.79 & 0.96 &   9 & 0.9502 & 28.827 \\
 28 & 355.43375 & -0.67539 & 0.2680 & 17.78 & 0.94 &  12 & 0.9107 & 43.606 \\
 29 & 357.93292 & 0.56078 & 0.2703 & 18.01 & 0.95 &  17 & 0.9083 & 72.542 \\
 30 & 357.23520 & -0.89120 & 0.2793 & 17.72 & 0.93 &   9 & 0.9034 & 59.979 \\
 \dag31 & 354.97913 & -0.43282 & 0.2769 & 17.53 & 0.94 &  13 & 0.8991 & 17.552 \\
 32 & 351.02548 & 0.36311 & 0.2780 & 17.90 & 0.93 &  12 & 0.8825 & 66.039 \\
 33 & 352.66733 & 0.60940 & 0.2757 & 17.66 & 0.92 &  14 & 0.8295 & 34.962 \\
 34 & 356.67987 & 0.83429 & 0.2637 & 17.89 & 0.93 &  15 & 0.8265 & 67.370 \\
 35 & 351.58939 & 0.16173 & 0.2870 & 18.15 & 0.88 &  12 & 0.7956 & 66.786 \\
 36 & 353.87994 & -0.88882 & 0.2974 & 17.95 & 0.93 &  11 & 0.7842 & 79.549 \\
 37 & 352.81964 & 0.72678 & 0.2812 & 18.34 & 0.93 &  11 & 0.7782 & 35.747 \\
 38 & 352.17755 & 0.82753 & 0.2784 & 17.42 & 0.92 &   9 & 0.7584 & 45.116 \\
 \dag39 & 353.87866 & -0.53116 & 0.2764 & 18.27 & 0.84 &  10 & 0.7452 & 18.973 \\
 40 & 356.76236 & -0.08544 & 0.2655 & 17.61 & 0.88 &  11 & 0.7327 & 63.065 \\
 41 & 358.12292 & 0.60623 & 0.2698 & 17.91 & 0.88 &  12 & 0.7018 & 76.620 \\
 42 & 355.63776 & -0.28874 & 0.2770 & 18.30 & 0.86 &  11 & 0.6924 & 26.124 \\
 43 & 358.60349 & 0.81066 & 0.2805 & 18.30 & 0.76 &  12 & 0.6835 & 83.396 \\
 44 & 351.32236 & 0.27741 & 0.2767 & 18.33 & 0.87 &   9 & 0.6553 & 60.071 \\
 45 & 351.95999 & -0.64258 & 0.2856 & 18.08 & 0.85 &   8 & 0.6424 & 60.422 \\
 46 & 358.44800 & 0.81641 & 0.2787 & 17.80 & 0.81 &   8 & 0.6040 & 79.441 \\
 47 & 355.18701 & 0.31234 & 0.2982 & 19.15 & 0.86 &  13 & 0.5994 & 79.790 \\
 48 & 357.44406 & 0.64598 & 0.2705 & 17.84 & 0.63 &   8 & 0.5988 & 63.779 \\
\enddata
\tablecomments{Data on 48 galaxy clusters listed  in
WHL catalog, within 90 Mpc comoving  distance from 
the center of Saraswati supercluster.  Abell 2631, the  most massive 
cluster and first in the list, is  considered as the center of the supercluster.  
The columns are as follows: (1) cluster number; (2,3) R.A. and decl. of the BCG; (4) redshift;
(5) r-band magnitude of the BCG;  (6,7,8) $R_{200}$, $N_{200}$ (richness parameter),  
and $M_{200}$; (9) Comoving distance between Abell 2631
and the cluster's  BCG. Clusters are arranged in decreasing order of their mass   
and those  marked with a \dag\  sign are the  five clusters within the bound core of the supercluster.}
\end{deluxetable*}

\begin{figure*}
\centering
\includegraphics[scale=0.27]{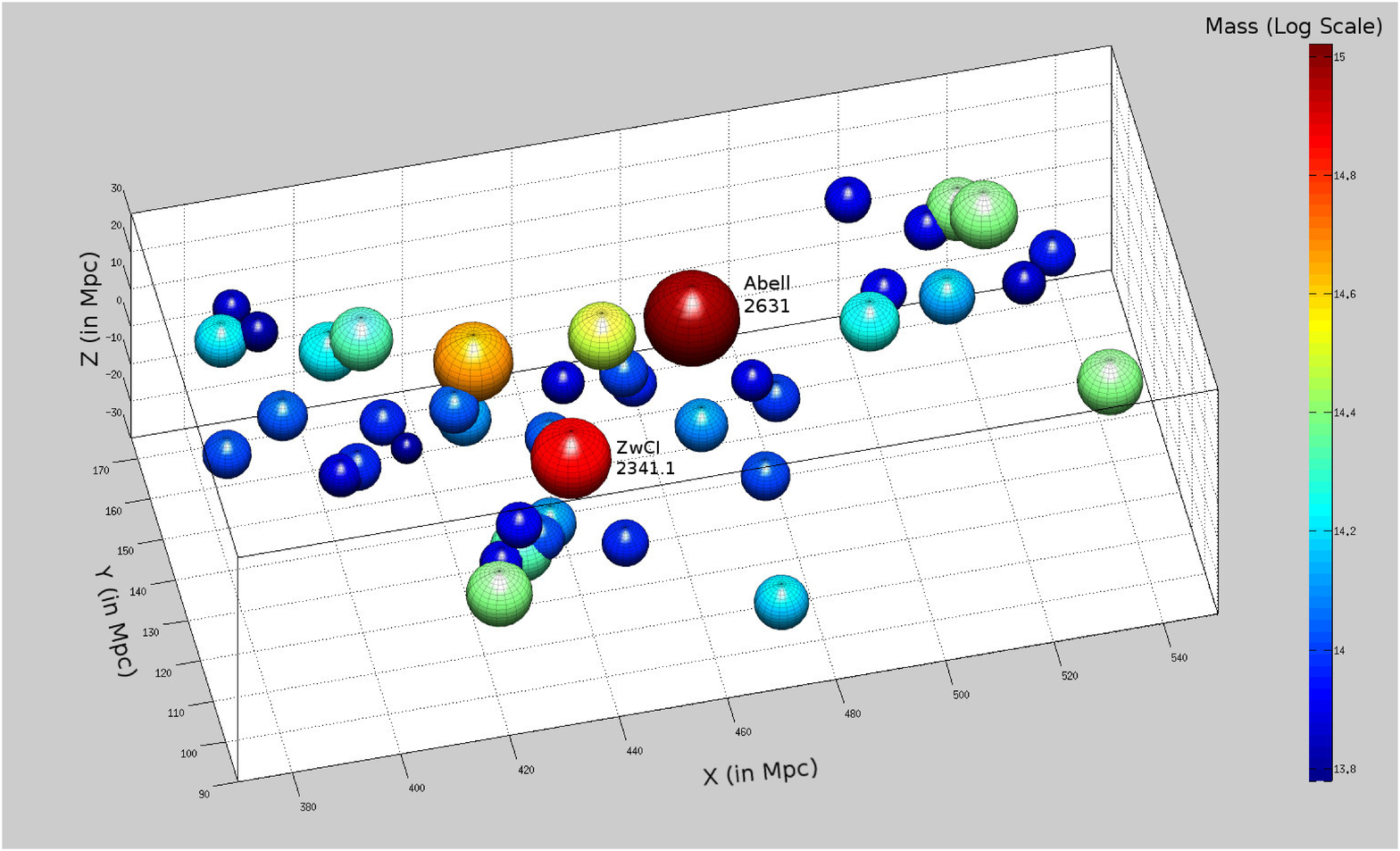}\\
\bigskip
\includegraphics[scale=0.7]{abc3.eps}
\caption{Upper: this figure shows the  3D distribution of the 
wall-like structure of the Saraswati supercluster,  mainly  comprising of
43 known clusters of galaxies,  here represented by spheres. The radius of each cluster  is 
proportional to its virial radius $r_{200}$. Colors represent the  masses of  clusters 
(in log($M_{200}$)) scale shown  with the color bar). 
The dark red  sphere represents  the  most  massive and rich galaxy cluster Abell 2631 and 
bright red the next  massive one,  ZwCl 2341.1+0000,  and  subsequently orange, lemon yellow, green, 
etc.  Dark blue spheres show the least
massive clusters.  Lower: 3D distribution of the  galaxies and some of the most massive clusters
 near the  Abell 2631  cluster are shown. Small blue  circles are  the  LBS galaxies, the
large golden circle represents  Abell 2631  and the other big red circles are the next four
most massive clusters of  Saraswati supercluster.\label{fig:saras}}
\end{figure*}

Figure \ref {fig:saras} (top) plots  the  3D  distribution (X,Y,Z comoving coordinates)  
of  43  clusters in Saraswati
supercluster,  spanning $\sim 200$ Mpc across,  and shows low-mass 
clusters surround the massive clusters like
irregular halos. In this figure the radius of a sphere (cluster) is proportional to its
$r_{200}$.  Colors represent the masses of the
clusters. The largest dark red sphere is  the most massive 
Abell 2631 ($M_{200}\approx 10^{15} M_{\odot}$; \cite{2012ApJ...75})  
and   small blue spheres show the  least massive clusters. Significantly, 
23  among these  43 clusters are very massive,  having  virial  
masses $M_{200} > 1 \times 10^{14} M_{\odot}$ (Table~\ref{table1}). 
This implies an unusually high-mass  concentration in  massive clusters 
within the supercluster region. In $\Lambda$CDM cosmology, the expected  
number of dark matter halos \citep{2001MNRAS.323....1S}
(with mass $M_{200} > 10^{14}M_{\odot}$)  within the Saraswati supercluster 
volume (estimated below), at  $z\sim 0.28$,  is only $\approx 2$.

\subsection{Bounded Region of the Saraswati--TurnAround Radius}

While a supercluster of the scale  of Saraswati
is unlikely to be in a state of  dynamical equilibrium with well-defined boundaries,
one can still ask whether or not  it is  gravitationally bound.
We now try to  estimate  how much  fraction of the  supercluster  is  possibly  gravitationally bound.

Although Saraswati is more planar than spherical, we
nevertheless apply the widely used spherical collapse model in which the turnaround radius
$R_{ta}$ (the radius of the shell separating infalling material from the outward expansion of the
Hubble flow \citep{1980lssu.book.....P,paddy00,2003ApJ...596..713B,2014JCAP...09..020P}
for a mass $M$ is

\begin{equation}
 R_{ta} = \left(\frac{3GM}{\Lambda c^{2}}\right)^{\frac{1}{3}},
\end{equation}
 or
\begin{equation}
  R_{ta} = \left(\frac{3GM}{H_{o}^2 \ \Omega_\Lambda}\right)^{\frac{1}{3}},
\label{Turnaround}
\end{equation}

where $\Lambda = (8 \pi G \rho_{\Lambda}/c^2)$  is the cosmological constant and 
  the dark energy density is $\rho_{\Lambda}$. Since the total halo mass of each cluster
  cannot be independently calculated, we use 
  results based on numerical simulations.
  As before,  total bound halo mass  of a component cluster is   taken to be  
  $M_{5.6} \approx 2.2 \times M_{200}$  
  following  \citep{2005MNRAS.363L..11B,Rines06,2013ApJ...767...15R},
  and thus  we  obtain   the sum of  masses of 43 clusters   as the    
  total   mass of  Saraswati, which  is   $M \approx  2 \times 10^{16} \ M_\odot$. 
  This is strictly a lower   limit because the mass of  matter (dark-matter + galaxies) 
  dispersed  within the intercluster space is not included. Below we have tried to correct for the
  extra mass.

\begin{figure*}
\centering
\includegraphics[scale=0.8]{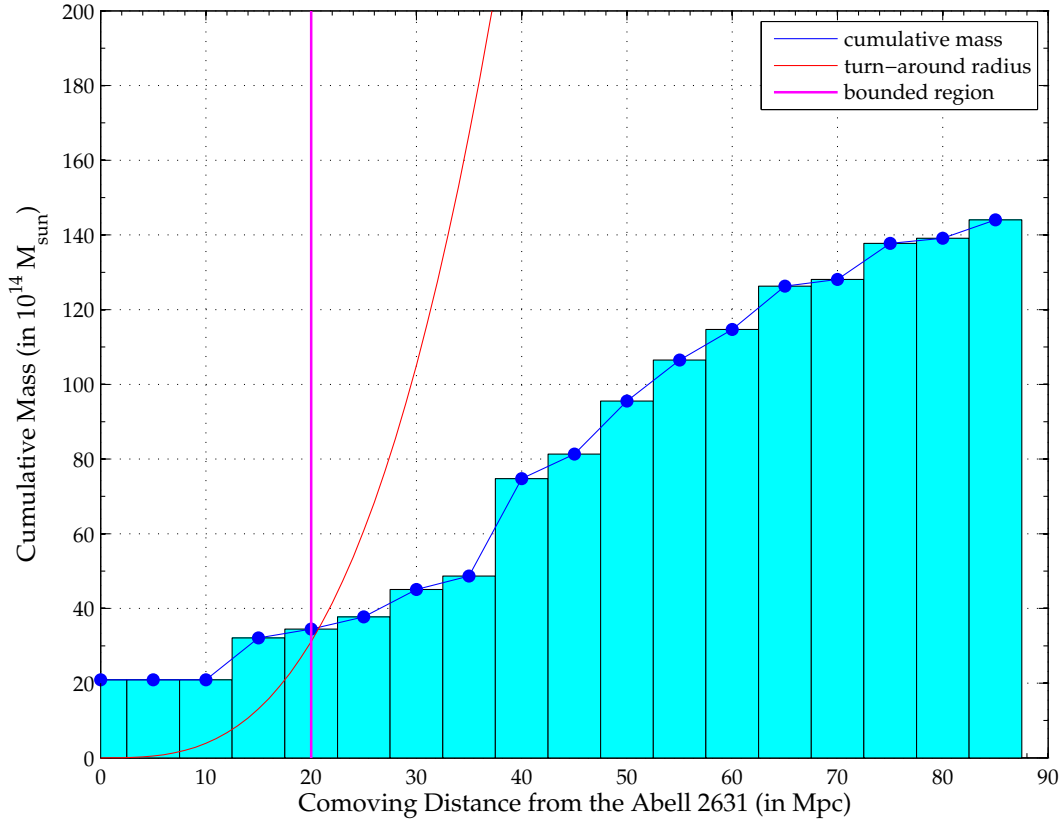}
\caption{The radial mass distribution of WHL clusters in the Saraswati  
supercluster up to  $R \approx 90$~Mpc radius.
The  x-axis is the radial  distance ($R$)  from  Abell 2631 (the most massive
galaxy cluster in the core of the Saraswati supercluster)  and the y-axis is the cumulative
bound halo mass of the  WHL clusters within the sphere of radius $R$  (bin size $\sim 5$ Mpc).
The red curve is the theoretical turnaround
radius (Equation~\ref{Turnaround}). The blue curve shows the cumulative mass distribution. 
Five  clusters  are inside the turnaround  curve and  38 are
outside it.  The clusters within the turnaround radius are  altogether bound
while the clusters beyond  it are unbound. The bounded region
has a comoving radius  of $\approx 20$ Mpc (magenta line).\label{fig:turn_around}}
\end{figure*}

Using this  total mass in equation  \ref{Turnaround} above, we 
obtain $R_{ta} \approx 30$~Mpc, or only about one-third of its $\approx 100$~Mpc physical radius. 
This suggests that the entire Saraswati supercluster cannot be a gravitationally 
bound  structure.  Further, by integrating in spherical
shells the enclosed  galaxy cluster masses up to a certain radius  $R$ and comparing that with  
 $R_{ta}$ (Figure \ref{fig:turn_around}), we infer 
 that only the mass within the radius  $R \sim 20$~Mpc is 
 possibly gravitationally
bound and  could have reached  turnaround.
Four high-mass galaxy clusters,  each  of  
mass  $M_{200} \sim   10^{14}$~$M_{\odot}$,
and  the  most  massive  Abell 2631  ($M_{200} \gtrsim   10^{15}$~$M_{\odot}$; \cite{2012ApJ...75}),
near the  center  comprise
this  bound core (Table~\ref{table1}), as  one would expect  from
the  gravitational collapse model.
The  aggregate  mass of this  dense,  bound  core
is at least $ 4 \times 10^{15} M_{\odot}$,  or about $  20\%$ of the  mass of the entire supercluster. 
Even neglecting the  (unknown)  small mass of  matter dispersed  within the intercluster space, 
this gives  a lower bound to the  core   overdensity  of  $ {\rho_{core}/\rho_{cz} = 1.4}$
and $\rho_{core}/\rho_{m} = 3.12$.
Here $\rho_{core}$, $\rho_{m}$ and $\rho_{cz}$ are the density of the bound core of Saraswati,
the mean matter density of the Universe at $z = 0.28$ and the critical density of the universe at $z = 0.28$, respectively.
According  to the  definition of a supercluster  proposed  by \cite{2015A&A...575L..14C},  
superclusters  are   structures  defined  based on an overdensity criterion that selects only those objects that will  
collapse in the future,  including those that are at a turnaround in the present epoch. Therefore, if we adopt this definition 
of a supercluster,  Saraswati   indeed qualifies for being  a  supercluster.

\section{Density Contrast of the Supercluster}

Here we calculate the density contrast of the Saraswati supercluster.
As we already have the halo mass information of all  the 43 clusters, we now need to
estimate the volume ($V$) of Saraswati to calculate its  mass  density. Since Saraswati is
a  wall-like  extended  structure  we do not assume a spherical volume 
because that would  largely overestimate $V$  and give a 
much smaller density contrast. Therefore, to obtain $V$  we calculate
 the  volume of the  Convex Hull  of the
43 clusters (set of points) within  Saraswati. A  Convex Hull of a set of points 
\textbf{P} in 3D Euclidean space that is 
the convex surface (envelope) of the minimum possible volume on \textbf{P}.
We use the MATLAB  {\em Qhull} algorithm \citep{Barber96} to construct the Convex Hull and to calculate 
its volume. In this way, we calculate the
comoving  volume of Saraswati as $V \approx  2.05 \times 10^{5}$ Mpc$^3$.
Now  the corresponding matter density contrast $\delta$ is

\begin{equation}
 \delta = \frac{\rho_{SS}}{{\rho_{m}}} - 1 = \left(\frac{\rho_{SS}}{\rho_{c}}  \frac{1}{\Omega_{M}}\right) - 1
\label{dens_contrast}
\end{equation},

where $\rho_{m}$ is the background matter density at redshift $z = 0.28$, $\rho_{c}$ is the
critical density at the present epoch,    and
$\rho_{SS}$ is the  comoving   mass density of the Saraswati supercluster. For the supercluster mass
we  use the total bound halo mass of  43 clusters,  estimated above to
be $\approx 2\times10^{16} M_\odot$, and use  that to calculate $\rho_{SS}$.
This gives an average   matter  density contrast  $\delta =  1.62$ ($\rho_{SS}/\rho_{m}  =2.62$)
and   $\rho_{SS}/\rho_{c} = 1.17$  for the  supercluster.
Both these numbers fall short of  the minimum  density  $\rho/\rho_{c} > 2.36$
and  $\rho/\rho_{m} > 7.86$ necessary  if the entire supercluster is to
remain   gravitationally bound at present, and also
in the distant future \citep{Nagamine03,Dunner06}. We point out that at the 
present epoch the largest known gravitationally bound, virialized
structures are the  galaxy clusters spanning  a  few Mpc across and  masses
$\sim  10^{14 - 15} M_{\odot}$.  If   the  space  between the clusters 
of the Saraswati  supercluster  is  uniformly filled 
with   matter  at  background  density $\rho_{m}$, then its  total
mass will be $\approx 4\times10^{16} M_\odot$,  giving a
higher density contrast $\delta \approx  3$  and $\rho_{SS}/\rho_{m} \approx 4$.

\section{Discussion and Conclusions}

Showing a significant  mass  overdensity   on  $\approx  200 $~Mpc scale,
the Saraswati supercluster  clearly stands out on the sky as an 
especially  rare,  and  possibly among the  largest
prominent density enhancements found  at the  medium-high redshift epochs
$0.2 < z < 0.6$. Only very  few such  massive superclusters are known at present.  
Observing  such massive,  large-scale   structures at early cosmic
aeons has major implications for cosmological studies, as the number  of  highest  density
peaks represented by clusters and  superclusters  and their growth function
places strong constraints on the nature of the
primordial fluctuation field and the dark energy equation of state,  which connects pressure ($P$)
to energy density ($\rho_{\lambda}$); $\mathrm {P = w(z)\, \rho_{\lambda}}$
 \citep{Sahni00,Frieman08,2011MNRAS.417.2938S,2012ApJ...759L...7P}. Further, to
ascertain if Einstein's general  relativity correctly  describes  the formation of  such
large structures, testing it  across a  wide range  of physical scales
and mass densities is necessary, and  superclusters like Saraswati  are  powerful  probes of
gravity in its  extreme large-scale, weak-field limit \citep{2010AnPhy.325.1479J}.
Numerical simulations are also very  important  to test and calibrate  the  competing cosmological
models  as applied to the  present structures \citep{2012ApJ...759L...7P}.

  Interestingly, the  redshift epoch  of the Saraswati complex  is  near
$\mathrm z = {(\Omega_{m}/\Omega_{\Lambda})^{1/3w} -1}$ $\approx 0.37$
(for $ w = -1$) when  energy density in matter ($\mathrm \Omega_{m}$) and dark
energy ($\mathrm \Omega_{\Lambda}$) became almost equal and the growth of
large-scale perturbations  virtually ceased due to the
accelerated expansion of the universe \citep{Frieman08}. This gives rise to  an important
question: Did the  major  growth of this supercluster take  place much
before the dark energy dominated era   ($\mathrm z \sim  0.7$),  or did it happen fairly recently,
when it entered into a quasi-linear  ($\mathrm \delta_{m} = O(0.1)$) growth  regime?
A  large-scale structure this massive
evolves   very slowly and therefore  it  may  reflect the whole history of galaxy formation
and  the  primordial  initial conditions that have `seeded' it.  The  mass and  spatial
scale  of the present superstructure is comparable  to that of the
Sloan Great Wall \citep{Gott05} ($\sim200 - 400$ Mpc)  and the
Shapley Concentration \citep{1989Natur.342..251R,2000MNRAS.312..540B,Proust06} ($\sim200$ Mpc),  
 two  of the largest  and  most massive structures found in the nearby  universe.  
 Other  important questions that arise from this
 work are as follows. By what  process have these  extremely large and  overdense cosmic structures formed
 and  is their existence is in accord with a homogeneous, isotropic universe with primordial
 Gaussian random phase density  fluctuations as predicted by the  inflation? 

The  physical mechanism of  supercluster formation is still not well understood. The probability of
 massive  Saraswati-like superclusters   arising from  tiny ($\delta(\rho)/\rho < 0.1$)
    perturbations  in the  primordial Gaussian random  density field is
 very  small \citep{paddy00}.  
Therefore,  significant  galaxy  concentrations developing
on the scale of  $ \sim 200 $ Mpc  or  more  are puzzling,  which might  challenge  the
belief   that the universe  becomes   (statistically)   homogeneous  and isotropic   on
scales $\gtrsim 100 \, h^{-1}$~ Mpc \citep{Sarkar09}.   
These massive  structures  provide  excellent model-independent tests of  general  relativity on the largest scales.
For example  \cite{2012ApJ...759L...7P} have  identified only a  few  such extreme-scale superclusters  (size$\sim 200 - 400$~Mpc)  in their
{\em Horizon Run 2}   simulation  covering  $7.2^{3} \, h^{-3}$~Gpc$^{3}$ volume. 
They conclude  that an initially homogeneous isotropic universe with primordial Gaussian 
random phase density fluctuations growing in accordance with the general relativity, can explain the 
richness and size of the observed large-scale structures in the SDSS.  More recently
 another  large,  galactic-wall-like structure  spanning  $\sim250$ Mpc  at  $z \sim 0.47$  was  identified in the
 BOSS survey \citep{2016A&A...588L...4L}. On the other hand,  
\cite{2011MNRAS.417.2938S} advocate that the formation of even a 
few  extremely large  and massive  structures like the 
Sloan Great Wall  in the local universe will result  in 
tension with the  Gaussian initial conditions.

Dark energy (either a  `cosmological constant' or some other form) 
appears to dominate the  present universe, which not only alters  its expansion rate 
but also affects the evolution of  perturbations  in the density of matter, 
slowing down the gravitational collapse of material in recent times \citep{Frieman08}. 
A powerful probe of  dark energy,  and how it affects  the time  
evolution of the gravitational potential of
superclusters and voids,  is  provided by   the  observation of 
secondary temperature anisotropies in  the 
cosmic microwave background (CMB)  imprinted  via the late-time 
integrated Sachs--Wolfe (and the Rees--Sciama effect in nonlinear regime; ISW+RS) effect
\citep{Nishizawa2014}.
The existence  of    large-scale  superclusters and  voids of $ \gtrsim 100$~Mpc size 
at  $z \sim 0.5$  is   implied   by  correlated  CMB temperature
fluctuations of  $\mid \Delta T \mid \approx 10 \pm 2 \mu K$ 
 ascribed  to  the late-time  ISW+RS   effect \citep{Granett08,Inoue10,2012MNRAS.426.2581G}. 
The significance  of  such observations is  still  under    scrutiny, 
but   if  confirmed  they  appear to be in   tension  with the   $\Lambda$CDM  
predictions, e.g., \citep{Nadathur12,2014ApJ...786..110C,2016ApJ...830L..19N}.
For the Saraswati supercluster we  try  to obtain  its   ISW+RS signal, using its  
matter  density contrast  $\delta =  1.62$ and effective 
comoving radius $r_c = [3 V/4 \pi]^{1/3} \approx 
40$ Mpc. In linear growth regime  we obtain   $\Delta T  \approx 1.6 \, \mu K$  on
$\approx 6.5 \deg$ scale,  which is  smaller than   some present 
 measurements \citep{Granett08,Inoue10,2012MNRAS.426.2581G}. However,  
 our   estimation  is     uncertain at present owing  to our lack of 
 detailed   knowledge about   the  shape  and  depth of 
 gravitational potential of this  supercluster  and  its  time evolution.  We highlight that finding  
   some more extremely  massive,  overdense   structures  like Saraswati,
the  Shapley  Concentration, and the Sloan/BOSS Great Wall at  higher  redshifts ($z \approx 0.3 - 1 $),  
 in  the dark energy dominated era,   and  measuring 
 their ISW+RS   and  Sunyaev--Zel'dovich  imprint  on the CMBR sky will  test  the  
 competing  cosmological  models and   provide  an  alternative  window  into  
 dark energy.

A  possible  clue to the formation of this  large galactic superstructure is provided
  by the detection of   huge   voids of $\sim40 - 170$~Mpc diameter observed
  around  the main  wall-like  overdense structure of Saraswati. 
  In $\Lambda$CDM cosmology, space  within  
  voids  (typically $\mathrm \delta(\rho)/{\bar \rho} \sim -0.8$;
  \cite{Colberg05})  expands faster  than the background Hubble expansion and thus
   matter  inside  voids  will  have  an  outward  component of  peculiar
   velocity  away from the void centers \citep{2005MNRAS.363..977P}. Hence,  matter
    within  voids   will  be   swept   up  into  dense sheets or pancakes,
   which intersect one another,  forming  long  galaxy filaments of the cosmic web
   separating  neighboring voids.  The overwhelmingly  large negative pressure  of  dark 
     energy  ($ P = w \,\, \rho_{\Lambda},  w = -1$  
     with  $ {\rho_{\Lambda}/\rho_{m}} \sim 10 - 20$ inside void)  
     and gravitational  attraction of the surrounding mass distribution  together   govern the 
  structure and  dynamics of  large-scale voids and possibly superclusters. 
   This  effect will be most pronounced  when  dark energy  becomes dominant over matter at  late times, around 
  $z \sim 0.37$ for  $w = -1$.  We believe that this  process   might have  played  an  important  role in the evolution of nearby   
      large-scale structures  like  Saraswati, the Shapley Supercluster, or the
      Sloan/BOSS Great Wall. Recently,   repulsion  from a pronounced void in
  galaxy distribution (the `dipole repeller') has been attributed to 
  the observed flow of nearby galaxies, converging  toward  a single attractor associated with the Shapley Supercluster \citep{Hoffman2017}, thus
  providing support to the model.

In the future, the Saraswati supercluster and its environs in   Stripe 82 region
 must be surveyed  in  greater depth  with
  more galaxy redshifts taken  on a  wider scale for a better  understanding of
   what physical processes were  involved  in  the
   growth of such enormous cosmic structures in  the  distant universe.  It also 
   offers exciting possibilities
   for gravitational weak-lensing mapping of dark matter distributed  in   clusters, filaments,
   and voids \citep{Vikram15}, and  for Sunyaev--Zel'dovich  imaging  of the  hot, baryonic
   matter  via its characteristic imprint on the cosmic microwave background photons.  This  
   region is  also  a prime target  for  future deep  surveys in  X-rays, from  the hot gas trapped in
 the numerous  overdense collapsing structures  (clusters) and from   the mass-accreting super massive black holes (active galactic nuclei),  and  
 for  deep radio observations  of  non-thermal  shocks  in  the  magnetized  cosmic web \citep{2016JApA...37...31K}. 
  This  supercluster  region  may contain large amounts  of warm-hot intergalactic medium (WHIM), which might be detected  
  via its spectral  signatures in soft X-rays and UV.  These  observations will open a  new window on the 
 study of large-scale structures in the universe.

\bigskip
\noindent
{\it Acknowledgement}

We thank the referee for the useful comments.
S.S. acknowledges support from a PhD studentship at IISER (Pune). J.B., J.J. and P.D. gratefully acknowledge generous
support from the Indo-French Center for the Promotion of Advanced Research
(Centre Franco-Indien pour la Promotion de la Recherche Avan\'{c}ee) under programme no. 5204-2.
J.J. acknowledges  generous support from IUCAA's  Associateship  Program. 
Funding for SDSS-III has been provided by the Alfred P. Sloan Foundation, the Participating Institutions,
the National Science Foundation, and the U.S. Department of Energy Office of Science. The SDSS-III Web site
is http://www.sdss3.org/. 
SDSS-III is managed by the Astrophysical Research Consortium for the Participating Institutions of the SDSS-III
Collaboration including the University of Arizona, the Brazilian Participation Group, Brookhaven National Laboratory,
Carnegie Mellon University, University of Florida, the French Participation Group, the German Participation Group,
Harvard University, the Instituto de Astrofisica de Canarias, the Michigan State/Notre Dame/JINA Participation Group,
Johns Hopkins University, Lawrence Berkeley National Laboratory, Max Planck Institute for Astrophysics,
Max Planck Institute for Extraterrestrial Physics, New Mexico State University, New York University, Ohio State University,
Pennsylvania State University, University of Portsmouth, Princeton University, the Spanish Participation Group, University of Tokyo,
University of Utah, Vanderbilt University, University of Virginia, University of Washington, and Yale University. 



\end{document}